\documentclass{ptephy}%%%%where ptephy is the template name

\usepackage{graphicx} %for dealing with figures.
\usepackage{url} %It provides better support for handling and breaking
\begin{document}

\title{Results from the Wilkinson Microwave Anisotropy Probe}

\author{
\name{Eiichiro Komatsu}{1,2} and  
\name{Charles L. Bennett}{3} 
(on behalf of the {\sl WMAP}
science team\thanks{The {\sl WMAP} science team includes C. Barnes,
R. Bean, C.~L. Bennett, O. Dor\'e, J. Dunkley, B. Gold, M.~R. Greason,
M. Halpern, R.~S. Hill, G. Hinshaw, N. Jarosik, A. Kogut, E. Komatsu,
D. Larson, M. Limon, S.~S. Meyer, M.~R. Nolta, N. Odegard L. Page,
H.~V. Peiris, K.~M. Smith, D.~N. Spergel, G.~S. Tucker, L. Verde,
J.~L. Weiland, E. Wollack, and E.~L. Wright.})} 
%%%%%%%%%%% The \name command should be used as \name{Insert author name here}{Insert affiliation number here}
%%%%% Please use \thanks for contributed author details

%%%%%%%%%%% The \affil command should be used as \affil{Insert affiliation number here}{Insert author address here}
\address{
\affil{1}{Max-Planck-Institut f\"ur Astrophysik,
Karl-Schwarzschild-Stra{\ss}e 1, 85740 Garching bei M\"unchen, Germany}
\affil{2}{Kavli Institute for the Physics and Mathematics of the
Universe, Todai Institutes for Advanced Study, the University of Tokyo,
Kashiwa, Japan 277-8583 (Kavli IPMU, WPI)}
\affil{3}{Department of Physics and Astronomy, Johns Hopkins University,
Baltimore, MD USA 21218}
\email{komatsu@mpa-garching.mpg.de}}

\begin{abstract}%
The {\sl Wilkinson Microwave Anisotropy Probe} ({\sl WMAP}) mapped
 the distribution of temperature and polarization over the entire sky in
 five microwave frequency bands.  These full-sky maps were used to 
 obtain measurements of temperature and polarization
 anisotropy of the cosmic microwave background with the unprecedented
 accuracy and precision. The analysis of two-point correlation functions of
 temperature and polarization data gives determinations of the
 fundamental cosmological parameters such as the age and composition of
 the universe, as well as the key parameters describing the physics of
 inflation, which is further constrained by three-point correlation
 functions. {\sl WMAP} observations alone reduced the flat
 $\Lambda$ cold dark matter ($\Lambda$CDM) cosmological model (six)
 parameter volume by a factor of  
 $>68,000$ compared with pre-{\sl WMAP} measurements.
 The {\sl WMAP} observations (sometimes in combination with
 other astrophysical probes) convincingly show the existence
 of non-baryonic dark matter, the cosmic neutrino background, 
 flatness of spatial geometry of the universe, a deviation from a 
 scale-invariant spectrum of initial scalar fluctuations, and that the
 current universe is undergoing an accelerated expansion. The {\sl WMAP}
 observations provide the strongest ever support for inflation; namely, the
 structures we see in the universe originate from quantum fluctuations
 generated during inflation.
\end{abstract}

\subjectindex{xxxx, xxx}

\maketitle

\section{Introduction}
The {\sl WMAP} \citep{bennett/etal:2003a} spacecraft was designed
to measure the full-sky distribution of temperature differences
(anisotropy) and polarization of the cosmic microwave background
(CMB). {\sl WMAP} is the successor of the legendary {\sl Cosmic Background
Explorer} ({\sl COBE}) satellite, whose spectrograph provided a precision-measurement
of the CMB blackbody, implying that matter and radiation were in thermal equilibrium, 
consistent with the expectation of the hot Big
Bang theory of the universe \citep{mather/etal:1990}.  The {\sl COBE} differential 
radiometers
discovered the primordial ripples in spacetime that existed in the early
universe \citep{smoot/etal:1992}. With 35 times better angular resolution
and 40 times better sensitivity than {\sl COBE}, {\sl WMAP} took the
cosmological research with CMB to the next level.

{\sl WMAP} was proposed to NASA as a MIDEX (Medium-Class Explorers)
mission in 1995. Four of eight Co-Investigators\footnote{The
Co-Investigators on the {\sl WMAP} proposal are C.~L. Bennett, G. Hinshaw,
N. Jarosik, S.~S. Meyer, L. Page, D.~N. Spergel, D.~T. Wilkinson (deceased), and
E.~L. Wright. Among them, Bennett, Hinshaw, Wilkinson, and Wright were
previously on the {\sl COBE} Science Team.} of  {\sl
WMAP} were previously on the {\sl COBE} Science Team. After being
selected in 1996, {\sl WMAP} launched on June 30, 2001, and arrived 
in its orbit around the second Lagrange point 
(L2), 1.5 million kilometers from Earth, three months later. Since then, 
{\sl WMAP} operated almost flawlessly for nine years until it
left its L2 orbit on September 8, 2010, to pass the baton to its successor,
the {\sl Planck} satellite, which arrived at L2 in July 2009.

The {\sl WMAP} team issued five data releases. The first-year data release 
(February 11, 2003) came with 13 papers
\citep{bennett/etal:2003b,bennett/etal:2003c,jarosik/etal:2003b,page/etal:2003a,page/etal:2003b,barnes/etal:2003,hinshaw/etal:2003a,hinshaw/etal:2003b,komatsu/etal:2003,kogut/etal:2003,spergel/etal:2003,verde/etal:2003,peiris/etal:2003}
and later with one more paper \citep{nolta/etal:2003};
the third-year data release (March 16, 2006) came with 4 papers
\citep{jarosik/etal:2007,hinshaw/etal:2007,page/etal:2007,spergel/etal:2007}
and later with one more paper \citep{kogut/etal:2007}; the five-year
data release (March 5, 2008) came with 7 papers
\citep{hinshaw/etal:2009,hill/etal:2009,gold/etal:2009,wright/etal:2009,nolta/etal:2009,dunkley/etal:2009a,komatsu/etal:2009}
and later with one more paper \citep{dunkley/etal:2009b}; the seven-year
data release (January 25, 2010) came with 6 papers
\citep{jarosik/etal:2011,gold/etal:2011,larson/etal:2011,bennett/etal:2011,komatsu/etal:2011,weiland/etal:2011};
and the final, nine-year data release (December 21, 2012) came with 2
papers \citep{bennett/etal:2013,hinshaw/etal:2013}. In addition, detailed
descriptions of the mission, data processing, calibration, as well as
of the data products for each release are given in the Explanatory 
Supplement document \citep{greason/etal:2012}.

In this article, we give a brief review on the {\sl WMAP} experiment,
the data analysis, and the main science results from the nine-year
observations.  

\section{How {\sl WMAP} measures temperature and polarization}

\subsection{Temperature}
\label{sec:temp}

{\sl WMAP} measured the distribution of temperature  and
polarization of the entire sky in five frequency bands (23, 33, 41, 61,
and 94~GHz). Five frequencies were necessary to separate the
CMB from emission of our own Galaxy, including synchrotron,
free-free, and dust emission (see section~\ref{sec:fg} for how to reduce
the effect of the Galactic emission).

{\sl WMAP} has two back-to-back mirrors to focus the incoming
electromagnetic waves arriving from two different lines of sight, which are
separated by 
$141^\circ$ from each other. The size of each mirror is 1.4~m$\times
1.6$~m, providing substantially better angular resolution than 
{\sl COBE}, which did not have mirrors but only horn antennas. The
incoming waves collected by two mirrors are received by a pair of
feed horns. Let us call these two inputs ``A side'' and ``B side.''

Each of the inputs from the A and B sides is separated into two orthogonal
polarized waves by Orthomode Transducers (OMTs), and then sent to a 
pair of radiometers. {\sl WMAP} has 20 radiometers, or 10 pairs of
radiometers. We call each pair of radiometers a ``differencing
assembly'' (DA). There 
are 1, 1, 2, 2, and 4 DAs at 23, 33, 41, 61, and 94~GHz,
respectively. Each DA measures temperature and polarization differences
between the A and B sides.

Each of 20 radiometers processes inputs from the A and B
sides as follows (section~2.3 of \citep{jarosik/etal:2003a})
\begin{itemize}
 \item [1.] The inputs are separated into two linear
combinations, $\frac1{\sqrt{2}}(A+B)$ and $\frac1{\sqrt{2}}(A-B)$.
 \item [2.] These linear combinations are amplified by High Electron 
 Mobility Transistor (HEMT) amplifiers, yielding
       \begin{eqnarray}
	u_1\equiv \left[\frac{A+B}{\sqrt{2}}+n_1\right]g_1,\qquad
        u_2\equiv \left[\frac{A-B}{\sqrt{2}}+n_2\right]g_2,
       \end{eqnarray}
where $n_1$ and $n_2$ are the noise added by the amplifiers, and $g_1$
       and $g_2$ are the gains of the amplifiers.
 \item [3.] The latter combination, $u_2$, is phased-switched to yield
       $\pm u_2$. Then $u_1$ and the phase-switched $u_2$ are combined
       to form $v_l\equiv u_1\pm u_2$ and $v_r\equiv u_1\mp u_2$.
 \item [4.] The combined signals are detected by the square-law detectors
       (diodes). The outputs, $V_l$ and $V_r$, are proportional to
       $v_l^2$ and $v_r^2$, respectively.
 \item [5.] Finally, we compute the difference between $V_l$  and $V_r$,
       obtaining 
\begin{equation}
 \frac{V_l-V_r}2=\frac{s}{2}(A^2-B^2)g_1g_2,
\end{equation}
where $s$ is the proportionality constant of the square-law
       detector. This quantity is thus proportional to the {\it difference}
       between the powers of light coming from the A and B sides; i.e.,
       {\sl WMAP} measures temperature differences in the sky
       separated by $141^\circ$, and the mean CMB temperature (2.725~K)
        and many types of undesirable systematic effects cancel out. 
\end{itemize}

We need to convert the measured $\frac12(V_l-V_r)$ (in units of
voltages) to a temperature difference in 
thermodynamic units. We do this by using the dipole anisotropy of the
CMB. As Earth orbits around Sun, the L2 point (hence {\sl WMAP})
also orbits around Sun at 30~km/s. This motion creates time-varying
dipole anisotropy, creating a sinusoidal signal changing over a year. As
we know the mean CMB temperature and the orbital velocity precisely, we
know that the amplitude of this signal must be $T_{\rm
cmb}v/c=273~\mu{\rm K}$. This fixes the proportionality constant between
$\frac12(V_l-V_r)$ and $T_A-T_B$, where $T_A$ and $T_B$ are the
temperatures toward the A and B sides, respectively. 

\subsection{Polarization}

\begin{figure}[t]
\centering\includegraphics[width=0.39\textwidth]{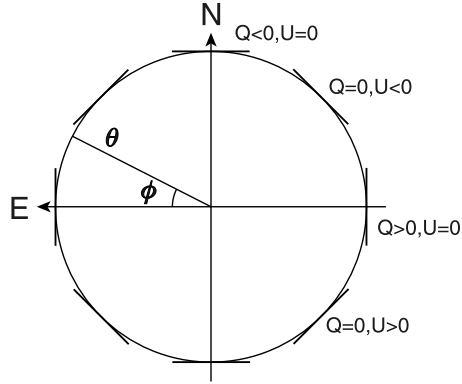}
\caption{Definition of the Stokes parameters with respect to 
 Galactic coordinates (adapted from \citep{komatsu/etal:2011}). ``N'' and
 ``E'' denote the Galactic north and east directions, respectively.}
\label{fig:stokes}
\end{figure}

The polarization information is still entangled in $T_A-T_B$ measured by
each radiometer. To measure polarization, we need to combine the
measurements of a pair of radiometers forming each DA. Let us define
$d_1\equiv (T_A-T_B)_1$ and $d_2\equiv (T_A-T_B)_2$, where $d_1$ and
$d_2$ are the outputs of two radiometers. The action of OMTs gives
(equations~10 and 11 of \citep{jarosik/etal:2007})
\begin{eqnarray}
\label{eq:d1}
 d_1&=& I_A+Q_A\cos 2\gamma_A+U_A\sin 2\gamma_A-I_B-Q_B\cos
  2\gamma_B-U_B\sin 2\gamma_B,\\
\label{eq:d2}
 d_2&=& I_A-Q_A\cos 2\gamma_A-U_A\sin 2\gamma_A-I_B+Q_B\cos
  2\gamma_B+U_B\sin 2\gamma_B.
\end{eqnarray}
Here, $I_A$, $Q_A$, and $U_A$ are the Stokes parameters describing the
incoming waves from the A side. We define the Stokes $Q$ and $U$ such
that the polarization directions of a pure $Q$ signal are parallel to
either Galactic longitudes or latitudes, and the polarization 
directions of  a pure $U$ signal are 45 degrees tilted from those of a pure $Q$
signal. In other words, a pure $Q$ signal
aligns with either the Galactic north-south or east-west
direction  (see figure~\ref{fig:stokes}). Then, $\gamma_A$ is the angle
between a meridian through the 
Galactic poles and the projection of the electric field of each output
port of the OMTs on the sky. 

The sum and difference of the instantaneous outputs of two radiometers
thus yield
\begin{eqnarray}
\label{eq:sum}
 \frac{d_1+d_2}2 &=& I_A-I_B,\\
\label{eq:diff}
 \frac{d_1-d_2}2 &=& Q_A\cos 2\gamma_A+U_A\sin 2\gamma_A-Q_B\cos
  2\gamma_B-U_B\sin 2\gamma_B.
\end{eqnarray}
The sum  gives the temperature difference (i.e., 
difference of unpolarized intensities), while the difference gives a
combination of the Stokes $Q$ and $U$.  

The polarization angles of radiometers were measured on the
ground using a polarized source. The uncertainty of the measurement is
$1^\circ$ (section 2.5 of \cite{page/etal:2003c}), and the measurements
are within $\pm 1.5^\circ$ of the design orientation (section 3 of
\cite{page/etal:2007}). In flight, we
observe Tau A \cite{page/etal:2007,weiland/etal:2011}, and find that the
standard deviation of angles measured in five bands is $0.6^\circ$ (see
Table 15 of \cite{weiland/etal:2011}). This is consistent with (and is
smaller than) the scatter of the ground measurements and we
conservatively use $1.5^\circ$ as an estimate of the systematic error in
the polarization angle of {\sl WMAP}. 

\subsection{Map making}

The instantaneous outputs of two radiometers per DA yield the
temperature and polarization differences between the A and B sides. The
next step is to reconstruct the distribution of temperature (minus the mean CMB
temperature) and polarization over the entire sky.

{\sl WMAP} scans the full sky in six months. We can then estimate 
maps of temperature and polarization over the full sky using the
six-month data. We first write down the measured time-ordered data (TOD)
as (section~3.4 of \citep{jarosik/etal:2007})
\begin{equation}
 d_t = \sum_p M_{tp}m_p + n_t,
\end{equation}
where $d_t$ is the TOD of two radiometers, $d=(d_1,d_2)$,
measured at a given observation time, $t$; $m_p$
is the actual sky map consisting of $m=(I,Q,U)$ at a given sky location
(pixel), $p$; $n_t$ is noise of the TOD; and $M_{tp}$ is the so-called ``mapping
matrix'' which projects $m_p$ onto $d_t$. 

For an ideal differential experiment, the mapping matrix is a $2N_t\times 3N_p$
matrix, where $N_t$ is the 
number of data-recording times and $N_p$ is the number of sky pixels. Each
row corresponds to one observation, while each column corresponds to a
map pixel. Each row of $M_{tp}$ has 6 non-zero elements.
The non-zero elements are $\pm 1$ in the columns
corresponding to the observed pixels in the $I$ map; 
and $\pm \cos 2\gamma_A$, $\pm \sin 2\gamma_A$, $\pm \cos 2\gamma_B$,
and $\pm \sin 2\gamma_B$ for the $Q$ and $U$ maps. The plus and minus signs are
chosen according to equations~(\ref{eq:d1}) and (\ref{eq:d2}). Each
observation is associated with 12 non-zero values of $M_{tp}$ that are
distributed in two rows.

In reality, there are two dominant nonidealities in radiometers that
must be taken into account. One is the ``bandpass mismatch.'' We take
the difference between two radiometers to measure polarization. While
these radiometers were designed and built to have nearly identical
frequency responses (bandpass) to the incoming electromagnetic waves, a
slight mismatch in the bandpass produces a spurious polarization
signal even in the absence of polarization. Suppose that the incoming
waves are unpolarized and have a spectrum of $I(\nu)$. Due to the
bandpass mismatch of two radiometers, they receive the incoming waves at
slightly different effective frequencies, $\nu_1$ and $\nu_2$. As a
result, the difference between two radiometers does not vanish,
producing a spurious polarization, $s$, given by
$s=I(\nu_1)-I(\nu_2)\approx (\nu_1-\nu_2)\partial I/\partial \nu$.
While the CMB, whose temperature does not depend on frequencies, does
not produce a spurious polarization, the other components (such as
Galactic emission) that depend on frequencies do produce a spurious
polarization. 

Fortunately, it is relatively straightforward to remove this
effect. Equations~(\ref{eq:d1}) and (\ref{eq:d2}) show that the real
polarization signals are modulated by the angle $\gamma$. On the other
hand, a spurious polarization is independent of $\gamma$. Therefore, we
can separate the real and spurious polarization signals if we have
enough coverage in $\gamma$. We modify equations~(\ref{eq:d1}) and
(\ref{eq:d2}) as (equations 17 and 18 of \citep{jarosik/etal:2007})
\begin{eqnarray}
\label{eq:d1s}
 d_1&=& I_A+Q_A\cos 2\gamma_A+U_A\sin 2\gamma_A+s_A-I_B-Q_B\cos
  2\gamma_B-U_B\sin 2\gamma_B-s_B,\\
\label{eq:d2s}
 d_2&=& I_A-Q_A\cos 2\gamma_A-U_A\sin 2\gamma_A-s_A-I_B+Q_B\cos
  2\gamma_B+U_B\sin 2\gamma_B+s_B,
\end{eqnarray}
and expand the mapping matrix to a $2N_t\times 4N_p$ matrix. (Each row
has 8 non-zero elements.) 
While {\sl WMAP}'s scan pattern allows for a uniform coverage in
$\gamma$ near the ecliptic poles, it covers only 30\% of possible
$\gamma$ on the ecliptic plane. This produces noisy modes in the
reconstructed sky maps, which must be properly de-weighted. (By comparison, 
{\sl Planck}'s coverage is $<4$\% on the ecliptic plane.)

The second non-ideality is the ``transmission imbalance,'' which is the
difference between the A and B sides; namely, the A and B sides do not
necessarily have equal responses to the incoming waves due to loss
(i.e., imperfect transmission) in the system. The spurious polarization
is an additive effect, but the transmission imbalance is a
multiplicative effect, given by (equations 19 and 20 of
\citep{jarosik/etal:2007}) 
\begin{eqnarray}
\nonumber
 d_1&=& (1+x_{im})[I_A+Q_A\cos 2\gamma_A+U_A\sin
 2\gamma_A+s_A]\\
\label{eq:d1ss}
& &-(1-x_{im})[I_B+Q_B\cos 2\gamma_B+U_B\sin 2\gamma_B+s_B],\\
\nonumber
 d_2&=& (1+x_{im})[I_A-Q_A\cos 2\gamma_A-U_A\sin
 2\gamma_A-s_A]\\
\label{eq:d2ss}
& &-(1-x_{im})[I_B-Q_B\cos 2\gamma_B-U_B\sin 2\gamma_B-s_B], 
\end{eqnarray}
where $x_{im}$ is the transmission imbalance factor, which has been
measured using the responses of radiometers to the CMB dipole
(table 2 of \citep{jarosik/etal:2007}). We include the transmission
imbalance in the 
mapping matrix by multiplying the A- and B-side elements by $1+x_{im}$
and $1-x_{im}$, respectively.

The optimal estimator for a sky map, $\tilde{m}_p$, that is unbiased and
has the minimum variance, is given by (in matrix notation)
\begin{equation}
 \tilde{m} = (M^TN^{-1}M)^{-1}M^TN^{-1}d,
\end{equation}
where $N$~$(=N_{tt'})$ is the noise matrix of the TOD. The TOD noise of
{\sl WMAP} is stationary, in a sense that the noise matrix is a
function of $\Delta t\equiv |t-t'|$. The  noise matrix is given by 
\begin{equation}
 N^{-1}(\Delta t)=\left(\begin{array}{cc}N_1^{-1}& 0\\ 0&
			   N_2^{-1}\end{array}\right),
\end{equation}
with (equation 21 of \citep{jarosik/etal:2007}) 
\begin{equation}
 N_i^{-1}(\Delta t)=\left\{
\begin{array}{ll}
C\left\{\int e^{i\omega\Delta t}[\int e^{i\omega
  t'}~N_i(t')dt']^{-1}d\omega+K\right\}, & \Delta t<\Delta t_{\rm max},\\
0,& \Delta t\ge \Delta t_{\rm max},
\end{array}
\right.
\end{equation}
where $\Delta t$ is in units of samples, and $\Delta t_{\rm max}$ is the
time lag at which the TOD noise correlation function ($N_i(t)$ in the
integral) crosses zero, typically $\approx 
600$~seconds. The coefficients, $C$ and $K$, are chosen such that
$N^{-1}_i(0)$ is normalized to unity, and that the mean of $N^{-1}_i(\Delta t)$
over $0\le \Delta t<\Delta t_{\rm max}$ vanishes. The TOD noise correlation
function, $N_i(t)$, is measured from the data and a functional form
is given in equation~(4) of \citep{jarosik/etal:2007}.
Given this noise matrix of the TOD, we solve the  equation,
$(M^TN^{-1}M)\tilde{m} = M^TN^{-1}d$, to find a sky map solution,
$\tilde{m}$, using the conjugate gradient method. 

\begin{figure}[t]
\centering\includegraphics[width=1\textwidth]{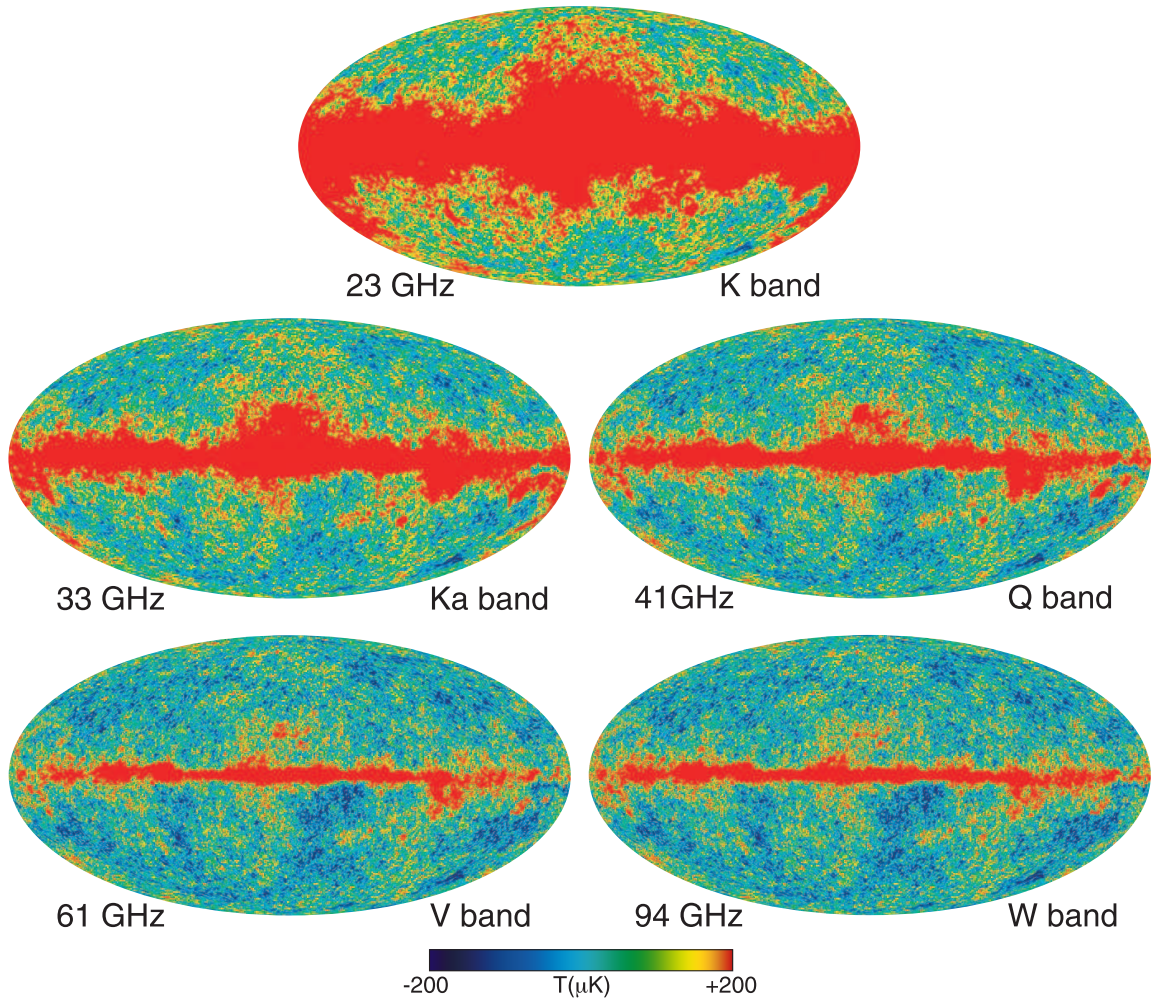}
\caption{Full-sky nine-year temperature maps in five frequency bands
 measured by  
 {\sl WMAP} (adapted from \citep{bennett/etal:2013}), shown in Galactic
 coordinates. Dipole anisotropy 
 has been removed from the maps. Maps are smoothed by a 0.2 degree
 Gaussian to suppress noise.}
\label{fig:wmap9_temp}
\end{figure}

\begin{figure}[t]
\centering\includegraphics[width=1\textwidth]{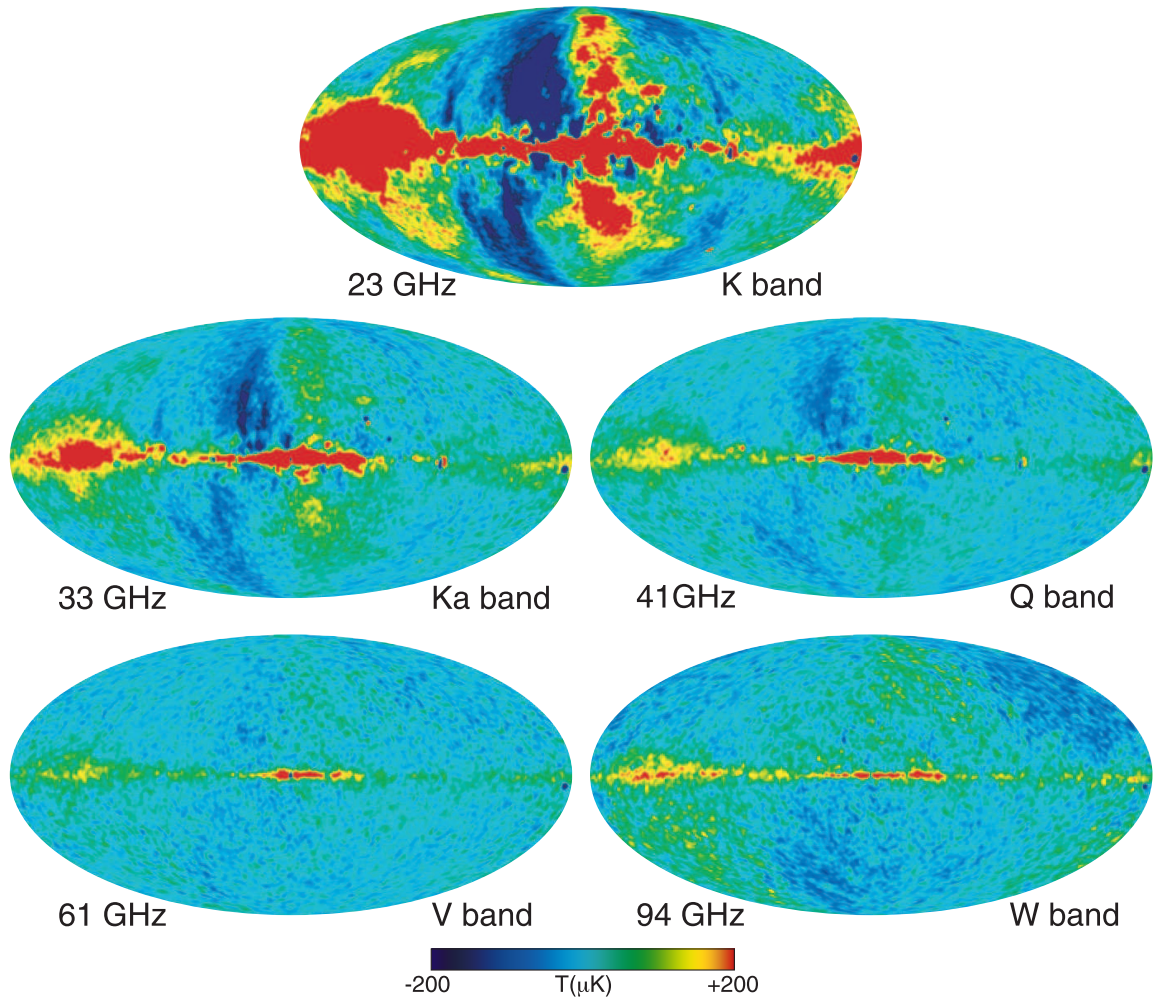}
\caption{Full-sky nine-year Stokes $Q$ maps in five frequency bands
 measured by  
 {\sl WMAP} (adapted from \citep{bennett/etal:2013}). Maps are smoothed
 to a common Gaussian beam of 2 degrees to suppress noise. See
 figure~\ref{fig:stokes} for the correspondence between the signs of $Q$
 and polarization directions.} 
\label{fig:wmap9_q}
\end{figure}

\begin{figure}[t]
\centering\includegraphics[width=1\textwidth]{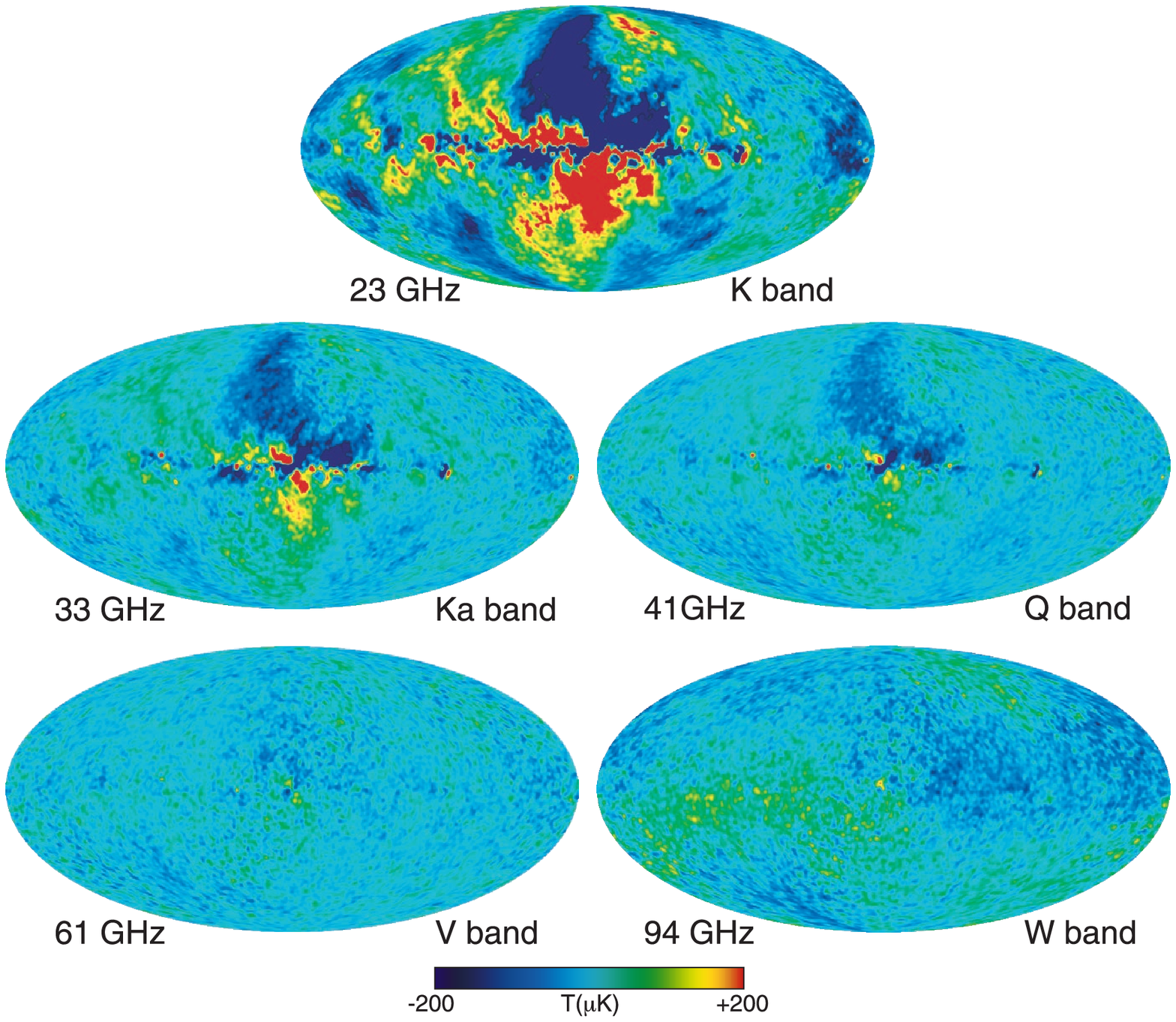}
\caption{Same as figure~\ref{fig:wmap9_u} but for Stokes $U$.}
\label{fig:wmap9_u}
\end{figure}

As described in section~\ref{sec:temp}, we use the CMB dipole to convert
the input signals (in voltages) to thermodynamic temperatures. The
uncertainty in this conversion (calibration uncertainty) per DA is 0.2\%
for the final nine-year temperature and polarization maps, which has
unchanged since the five-year analysis (section~4 of
\citep{hinshaw/etal:2009}). 
Figure~\ref{fig:wmap9_temp} shows the nine-year solutions to the full-sky
$I$ maps (minus dipole anisotropy), while
figures~\ref{fig:wmap9_q} and \ref{fig:wmap9_u} show the Stokes $Q$ and
$U$ maps, respectively, in five frequency bands.


\section{Galactic and extra-galactic foreground emission}
\label{sec:fg}
\subsection{Temperature}
Figure~\ref{fig:wmap9_temp} shows that the distribution of the measured
temperatures at high frequencies (41, 61, and 94~GHz) at high Galactic
latitudes are quite similar. This means that, at these frequencies, the
temperature data at high Galactic latitudes are dominated by the CMB
which is independent of frequencies. On the other hand, the data at
lower frequencies (23 and 33~GHz) are clearly affected by strong
Galactic emission, and the data near the Galactic plane are
dominated by the Galactic emission at all five frequencies.
Also, there are many extra-galactic sources (most of which are
synchrotron sources) all over the sky, which need to be removed from the
cosmological analysis.

To reduce the effects of the so-called ``foreground'' emission from our
Galaxy and extra-galactic sources, we combine two methods: one is to
simply mask the pixels which are strongly affected by the foreground
emission; and the other is to estimate and remove the foreground
emission from the sky maps. 

Three Galactic foreground components  are known to dominate in the
{\sl WMAP} frequencies (section~3 of \citep{bennett/etal:2003c}): synchrotron,
free-free, and dust. The antenna 
temperatures of these three components go (very approximately) as
$\propto \nu^{-3}$, $\nu^{-2}$, and $\nu^2$, respectively. As a result,
synchrotron dominates at the lowest frequencies (23 and 33~GHz),
free-free dominates in some regions in the sky at the intermediate
frequencies (41 and 61~GHz), and dust dominates at the highest frequency
(94~GHz). 

We define the mask for the temperature data as follows
(section~2 of \citep{gold/etal:2009}). We first smooth 
the 23~GHz (K band) map to one degree resolution, and remove an estimate
of the CMB from this map. The CMB is estimated by the internal
linear combination (ILC) method (section~5.2 of
\citep{hinshaw/etal:2007}). We then mask 
the pixels brighter than a certain threshold temperature, until 75\% or
85\% of the sky is left unmasked. We repeat the same procedure for the
41~GHz (Q band) map. The masks defined by the K- and Q-band maps are
added to form two masks, ``KQ75'' and ``KQ85'' masks, depending on how
much sky was left unmasked in the K- or Q-band map. 

The mask for extra-galactic sources is created using the locations of the known
bright radio sources in the literature (as described in section~7 of
\citep{bennett/etal:2003c}), the 
 sources found in the {\sl WMAP} nine-year data (section~5.2.2 of
 \citep{bennett/etal:2013}), and 
 additional sources found in the {\sl 
Planck} early release compact source catalog at 100~GHz
\citep{planck_early_vii:2011}. 
An exclusion radius of 1.2 degree is used for
sources brighter than 5~Jy in any {\sl WMAP} band, and an exclusion
radius of 0.6 degree is used for fainter sources.

To further reduce foreground emission in unmasked pixels, we
estimate the distribution of the diffuse Galactic foreground emission
over the full sky and remove it from the {\sl WMAP} maps. We use the
difference between 
the 23 and 33~GHz maps (which does not contain the CMB) for synchrotron
emission; a map of H$\alpha$ \citep{finkbeiner:2003} corrected for
extinction and scattering (section~5.3.1 of \citep{bennett/etal:2013})
for free-free emission; and a map of dust emission
\citep{finkbeiner/davis/schlegel:1999}. These three maps are
simultaneously fit to and removed from the 41, 61, and 94~GHz maps,
yielding the foreground-reduced maps at these frequencies.

Finally, we slightly enlarge the KQ75 and KQ85 masks by further masking
the regions which have significant excess in the differences between the
foreground-reduced 41 and 61~GHz maps, and 61 and 94~GHz maps.
The resulting nine-year temperature masks, ``KQ75y9'' and ``KQ85y9,''
retain 68.8\% and 74.8\% of the sky, respectively.
The former is used for testing Gaussianity
of the temperature data, while the latter is used for the power spectrum
analysis. 

\subsection{Polarization}
Figures~\ref{fig:wmap9_q} and \ref{fig:wmap9_u} show that the
polarization maps at low frequencies are dominated by the Galactic
emission. We thus mask the regions strongly contaminated by foreground
emission, and remove an estimate of the foreground emission from 
unmasked pixels.

Two polarized foreground components are known to dominate in the {\sl
WMAP} frequencies \citep{page/etal:2007}: synchrotron and
dust. Therefore, one might think that 
reducing foreground in polarization is easier than in temperature, as we
have fewer foreground components. In reality, the polarization analysis
is more challenging because the CMB signal in polarization is 10 times fainter
than in temperature. While the polarized foreground emission is clearly
seen in figures~\ref{fig:wmap9_q} and
\ref{fig:wmap9_u}, there is no clear evidence for the CMB signals; thus,
we need more sophisticated statistical analysis to extract faint
polarization signals of the CMB.

We define the mask of polarization maps using the 23~GHz map (which is
dominated by synchrotron) and a model of dust emission
\citep{page/etal:2007}. We first create 
lower resolution $Q$ and $U$ maps at the 23~GHz, each containing 3072
pixels. We then compute the total polarization intensity,
$\sqrt{Q^2+U^2}$. As the presence of noise produces a small positive bias in
this quantity, we remove it using our estimate of noise in the
maps. Finally, we mask the pixels that are brighter than a certain
threshold polarization intensity. We choose the threshold to be 0.6
times the mean polarization intensity, and call this mask the ``P06'' mask.
As for dust, we use a map of dust emission constructed from the {\sl
WMAP} temperature data using Maximum Entropy Method (MEM)
(section~5 of \citep{bennett/etal:2003c}). We choose a threshold in this
map to be 0.5 
times the maximum value found in the polar caps ($|b|>60$ degrees), and
mask the pixels brighter than this threshold value. We then add this
dust mask to the P06 mask. We find that extra-galactic sources are
minimally polarized in the {\sl WMAP} frequencies; thus, we only mask
ten bright sources outside of the P06 mask
\citep{page/etal:2007,bennett/etal:2013}. The combined P06 mask including
synchrotron, dust, and extra-galactic sources retains 73.2\% of the sky
for the cosmological analysis. 

To further reduce foreground emission in unmasked pixels, we use
the 23~GHz polarization maps to trace synchrotron. As for dust, we use
(equation~15 of \citep{page/etal:2007})
\begin{eqnarray}
 Q_{\rm dust}(\hat{n})&=&I_{\rm dust}(\hat{n})g_{\rm
  dust}(\hat{n})\cos[2\gamma_{\rm dust}(\hat{n})],\\
 U_{\rm dust}(\hat{n})&=&I_{\rm dust}(\hat{n})g_{\rm
  dust}(\hat{n})\sin[2\gamma_{\rm dust}(\hat{n})],
\end{eqnarray}
where $I_{\rm dust}$ is the same dust map that we used for the temperature
analysis \citep{finkbeiner/davis/schlegel:1999}.
Dust emission is polarized because dust grains are not spherical, and
are aligned with coherent magnetic fields in our Galaxy such that the
semi-major axes of grains are perpendicular to the fields. As a result, the
polarization directions of dust emission are perpendicular to the field
directions.  The dust
polarization we observe along a particular line of sight is the
projection of multiple polarization signals along the line of sight,
which is affected by geometry of the fields. We take this
projection effect into account by the function $g_{\rm dust}$, which is
calculated using a simple model of fields given in section~4.1 of 
\citep{page/etal:2007}. 

How do we estimate $\gamma_{\rm dust}$? As the polarization directions
of synchrotron are also perpendicular to the field directions, we could
take the polarization directions at 23~GHz as an estimate for
$\gamma_{\rm dust}$; however, alignment of dust grains with the fields
is not necessarily perfect, which yields some differences between
$\gamma_{\rm dust}$ and $\gamma_{\rm synch}$. Therefore, we use the
polarization directions measured toward stars. While the intrinsic starlight is
usually unpolarized (or polarized very weakly), the observed starlight
can be polarized due to selective extinction of the starlight when it
passes through regions with dust grains. As it gets more extinction along the semi-major
axes of dust grains, the observed starlight polarization direction is
precisely orthogonal to the polarization direction of dust emission from
the same location. We have compiled the existing measurements of 
starlight polarization in the literature (section~4.1.2 of
\citep{page/etal:2007}), and created a map of the starlight polarization
directions, $\gamma_*(\hat{n})$. We then compute the dust polarization
direction as $\gamma_{\rm dust}(\hat{n})=\gamma_*(\hat{n})+\pi/2$.
The correlation between $\gamma_{\rm dust}$ computed in this way and
$\gamma_{\rm synch}$ shows that these two angles typically agree to 20 degrees.

We simultaneously fit the 23~GHz polarization maps and dust polarization
maps to reduce polarized foregrounds from the higher frequency maps at 33, 41, 61, and
94~GHz. Due to much lower signal-to-noise ratios in the polarization
maps, removal of the polarized foreground presents a great challenge to the
{\sl WMAP} analysis. The error in the foreground removal has a
non-negligible impact on the inferred value of the optical depth to the
Thomson scattering, $\tau$. For example, an alternative foreground
removal method presented in \citep{dunkley/etal:2009b} shifts the value
of $\tau$ as much as the 1-$\sigma$ statistical uncertainty.

\section{Power spectrum measurements}
The steps we have described so far give us foreground-reduced
temperature maps at 41, 61, and 94~GHz, and foreground-reduced
polarization maps at 33, 41, 61, and 94~GHz. The temperature maps come
with the KQ75 and KQ85 masks depending on the purpose of the
cosmological analysis, and the polarization maps come with the P06
mask. In this section, we describe how to measure the angular power
spectra from these maps.
\subsection{Temperature}

Assuming that the distribution of CMB temperatures in the foreground-reduced
maps outside the mask is given by a Gaussian distribution, which is a
good approximation as described in section~\ref{sec:ng}, a complete
description of the measured CMB temperatures is given by the
following probability density distribution function (PDF):
\begin{equation}
 p(T) = \frac{\exp\left[-\frac12\sum_{ij}\sum_{ab}\delta
			       T_i^{(a)}(S+N)^{-1}_{iajb}\delta
			       T_j^{(b)}\right]}{\sqrt{\det[2\pi(S+N)]}},
\label{eq:pdf}
\end{equation}
where $\delta T_i^{(a)}\equiv T_i^{(a)}-T_{\rm cmb}$ is the difference
between the CMB temperature toward a direction (or a sky pixel) $i$ and
the mean CMB temperature, $T_{\rm cmb}=2.725$~K, with an extra index $a$
denoting a DA and an observed period (e.g., first year,
second year, etc). The total covariance matrix consists of the signal
matrix, $S_{iajb}$, and the noise matrix, $N_{iajb}$. These matrices are 
$M_{\rm pix}M_{\rm DA}M_{\rm year}\times M_{\rm pix}M_{\rm DA}M_{\rm
year}$ matrices, where $M_{\rm pix}$ is the number of pixels, $M_{\rm
DA}=6$ is the number of DAs used for the temperature analysis (2 at
61~GHz and 4 at 94~GHz), and $M_{\rm year}=9$ is the number of years.
For {\sl WMAP}, the noise
matrix vanishes unless $a=b$, i.e., noise is uncorrelated between different
DAs or years. 

As {\sl WMAP} is a differential experiment measuring temperature
differences between two points separated by 141 degrees, there is a
pixel-to-pixel noise correlation at 141 degrees (figure~11 of
\citep{hinshaw/etal:2003a}). However, this correlation has a negligible
influence on the temperature analysis, as 
the signal covariance due to the CMB totally dominates at large angular
scales where this pixel correlation is important. We thus have a simple
description of the noise matrix:
\begin{equation}
 N_{iajb}=\frac{[\sigma_{0}^{(a)}]^2}{n_{{\rm obs},i}^{(a)}}\delta_{ij}\delta_{ab},
\end{equation}
where $\sigma_0^{(a)}$ sets the overall noise level per DA per
observation, and $n_{{\rm obs},i}^{(a)}$ gives the corresponding
effective number of observations in a sky pixel $i$.
We also set the noise matrix to have a large value in the masked pixels,
i.e., we effectively set the noise level to be infinity at the masked pixels.

The CMB signal matrix is given by
\begin{equation}
 S_{iajb} = \frac1{4\pi}\sum_\ell (2\ell+1)C_\ell b_\ell^{(a)}b_\ell^{(b)}P_\ell(\cos\theta_{ij}),
\end{equation}
where $C_\ell$ is the temperature power spectrum of the CMB, and
$b_\ell^{(a)}$ is the so-called ``beam transfer function,'' which is the
Legendre transform of a symmetrized beam profile of a given DA, and
$P_\ell(x)$ is the usual Legendre polynomials. An additional smearing
due to pixelization may also be included in $b_\ell^{(a)}$.

The beam transfer functions of all DAs are determined by the repeated 
observations of Jupiter and the full physical modeling of the optical system
of {\sl WMAP} \citep{page/etal:2003a,hill/etal:2009}. Two Jupiter
observing seasons of $\sim 50$ days each occur every 395--400 days, and
we had 17 seasons of the Jupiter data over nine years of operation.
The Jupiter data are used to determine the inner beam profiles directly
out to radii at which the antenna gains drop to 2, 3, 5, 6, and 9 dBi
(or $-45.0$, $-46.4$, $-46.3$, $-48.8$, and $-48.8$~dB) at 23, 33, 41,
61, and 94~GHz, respectively (section~3 of
\citep{bennett/etal:2013}). Then, the physical optics modeling is used 
from the inner beams out to 7.0, 5.5, 5.0, 4.0, and 3.5 degrees from the
beam center at 23, 33, 41, 61, and 94~GHz, respectively. 
The beam response at greater distances from the beam center (i.e., far
sidelobes) has been measured on the ground before launch and in-flight
using Moon \citep{barnes/etal:2003}.

The accurate determination of the beam transfer functions is crucial for
the accurate recovery of the intrinsic CMB power spectrum,
$C_\ell$, as the errors we make in $b_\ell^{(a)}$ propagate
directly into $C_\ell$. We estimate that the 1-$\sigma$ uncertainty
in the recovered $C_\ell$ from the nine-year data due to the
uncertainty in the beam transfer functions is 0.6\% at $\ell\gtrsim
100$. This uncertainty is coherent (correlated) over a wide range in
$\ell$, and must be included in the cosmological analysis of the CMB
power spectrum. 

How do we infer cosmological parameters from the CMB maps? 
We can  calculate $C_\ell$ as a function of cosmological parameters
using the linear Boltzmann code such as {\sf CMBFAST}
\citep{seljak/zaldarriaga:1996}, {\sf CAMB}
\citep{lewis/challinor/lasenby:2000}, and {\sf CLASS}
\citep{blas/lesgourgues/tram:2011}. Ideally, we wish to 
evaluate equation~(\ref{eq:pdf}) directly as a function of cosmological
parameters given our knowledge of noise, mask, and beam transfer
functions; namely, we interpret equation~(\ref{eq:pdf}) as the
likelihood function of temperature data given cosmological
parameters, $p(T|\theta)$, where $\theta$ denotes a set of cosmological
parameters. We then use Bayes' theorem to obtain the posterior
probability  of the parameters given the temperature data as
$p(\theta|T) \propto p(T|\theta)p(\theta)$. Here, $p(\theta)$ is the
prior probability  of cosmological parameters, and we take it to be
uniform within a certain reasonable range of $\theta$. We then calculate the
best-fit values of $\theta$ and the 68\% confidence intervals, etc.

Another approach is to use equation~(\ref{eq:pdf}) as the
likelihood function of temperature data given power spectra,
$p(T|C_\ell)$, and obtain the posterior probability as 
$p(C_\ell|T) \propto p(T|C_\ell)p(C_\ell)$ with uniform $p(C_\ell)$
within a certain reasonable range of $C_\ell$. We then evaluate $p(C_\ell|T)$
for theoretically computed $C_\ell$ with various values of $\theta$, and
find the best-fit values of $\theta$ and the 68\% confidence intervals,
etc. In other words, we write the posterior probability of $\theta$ as
$p(\theta|T) \propto p(T|\theta)p(\theta) = \int dC_\ell~p(T|C_\ell)p(C_\ell|\theta)p(\theta)$, where
$p(C_\ell|\theta)=\delta[C_\ell-C^{\rm theory}_\ell(\theta)]$ because we
know how to calculate $C_\ell$ theoretically as a function of $\theta$.

The difference between these two approaches is that the latter approach
produces a convenient intermediate product, $p(C_\ell|T)$, which can be
made much faster to evaluate than the full likelihood function using
the so-called ``Blackwell-Rao (BR) estimator'' \citep{wandelt/etal:2004}.
Note that the form of $p(C_\ell|T)$ is non-Gaussian even though
$p(T|C_\ell)$ is a Gaussian (equation~\ref{eq:pdf}), as $C_\ell$ is
a quadratic function of temperatures. The central limit theorem makes
$p(C_\ell|T)$ closer to a Gaussian distribution for large values of
$\ell$, but it is important to use the full non-Gaussian form of
$p(C_\ell|T)$ at small values of $\ell$ for the cosmological parameter
estimation. 

While it is certainly possible to obtain $p(C_\ell|T)$ using
the BR estimator out to large values of $\ell$
\citep{o'dwyer/etal:2004,eriksen/etal:2004}, the computational cost is
still quite substantial. Therefore, the {\sl WMAP} team has adopted a
hybrid approach: we use the BR estimator to compute the full
$p(C_\ell|T)$ at $\ell\le 32$ from the ILC map with the KQ85 mask
(following section~6 of \citep{bennett/etal:2013}), and use the so-called
``quadratic maximum likelihood (QML) estimator''
\citep{tegmark:1997,bond/jaffe/knox:1998} for $\ell>32$. The QML
estimator, $\hat{C}_\ell$, is obtained  by Taylor-expanding the
logarithm of equation~(\ref{eq:pdf}) up to second order in
$\hat{C}_\ell-C_\ell$, and maximizing it with respect to $C_\ell$, i.e.,
$\left. d\ln p(T|C_\ell)/dC_\ell\right|_{\hat{C}_\ell}=0$. The solution is 
\begin{equation}
 \hat{C}_\ell = \sum_{\ell'} (F^{-1})_{\ell \ell'}
  \frac1{2\ell'+1}\sum_{ab}\sum_{m'} \tilde{a}_{\ell' m'}^{(a)}\tilde{a}_{\ell' m'}^{(b)*},
\end{equation}
where $\tilde{a}_{\ell m}^{(a)}$ is the spherical harmonics coefficient
of a map filtered by $(S+N)^{-1}$, $\tilde{a}_{\ell m}^{(a)}\equiv \int
d^2\hat{n}~Y_{\ell m}^*(\hat{n})[(S+N)^{-1}\delta T]^{(a)}(\hat{n})$. The
matrix $F_{\ell \ell'}$ is given by
\begin{eqnarray}
\nonumber
 F_{\ell\ell'} &\equiv& -\left<\frac{\partial^2\ln p}{\partial
  C_\ell\partial C_{\ell'}}\right>\\
\nonumber
&=& \frac{(2\ell+1)(2\ell'+1)}{2(4\pi)^2}
(S+N)^{-1}_{j'b'ia}b_\ell^{(a)}P_\ell(\cos\theta_{ij})b_{\ell}^{(b)}(S+N)^{-1}_{jbi'a'}b_{\ell'}^{(a')}P_{\ell'}(\cos\theta_{i'j'})b_{\ell'}^{(b')},\\
\end{eqnarray}
where the repeated indices are summed.

The QML estimator gives our best estimate for $C_\ell$ at each $\ell$ if
we use the correct $C_\ell$ in the $S$ matrix. We can therefore improve the
performance of the estimator by iterating the estimation: assume some
reasonable $C_\ell$, estimate $C_\ell$, use the estimated $C_\ell$ to
recompute the QML estimator, and repeat. If an incorrect $C_\ell$ is
used, the QML estimator does not give the minimum variance, but it is
still unbiased.

While the QML estimator gives an estimate, we still need to calculate the
form of the posterior distribution of $C_\ell$. We know 
that a Gaussian approximation is not accurate enough; thus, we combine a
Gaussian distribution and a log-normal distribution with appropriate
weights to obtain an improved form of the posterior distribution
(following section~2.1 of \citep{verde/etal:2003}). We use the nine-year
temperature data at 61 and 94~GHz, which have the highest angular
resolutions, to compute $C_\ell$.

\begin{figure}[t]
\centering\includegraphics[width=0.9\textwidth]{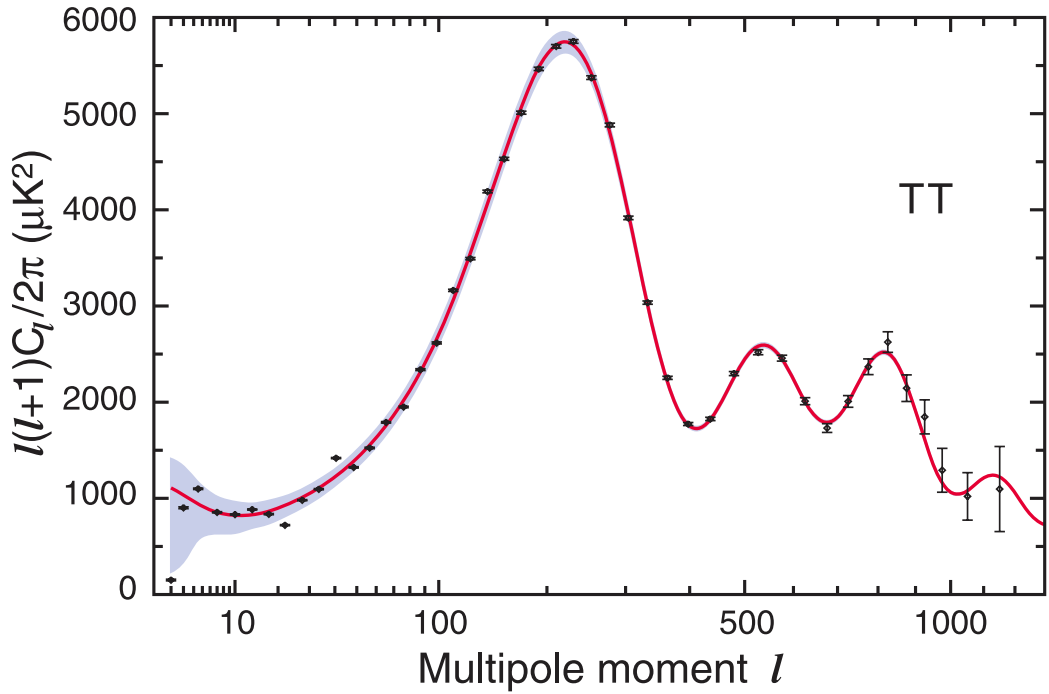}
\caption{Nine-year angular power spectrum of the CMB temperature
 (adapted from \citep{bennett/etal:2013}). While we measure $C_\ell$ at
 each $\ell$ in $2\le \ell\le 1200$, the points with error bars show
 the binned values of $C_\ell$ for clarity. The error bars show the 
 standard deviation of $C_\ell$ from  instrumental noise, $[2(2C_\ell
 N_\ell+N_\ell^2)/(2\ell+1)f_{{\rm 
 sky},\ell}^2]^{1/2}$. The shaded area shows the standard deviation from the
 cosmic variance term, $[2C_\ell^2/(2\ell+1)f_{{\rm
 sky},\ell}^2]^{1/2}$ (except at very low $\ell$ where the 68\%~CL from
 the full non-Gaussian posterior probability is shown). The solid line
 shows the theoretical curve of the best-fit $\Lambda$CDM cosmological model.}
\label{fig:cltt}
\end{figure}

Figure~\ref{fig:cltt} shows the nine-year measurements of $C_\ell$ along
with estimates of the 68\%~CL error bars. The error bars are calculated
as follows. Given the form of
$p(C_\ell|T)$, we calculate the second-order moment (variance) of
$C_\ell$, and parametrize it as 
\begin{eqnarray}
\label{eq:variance}
\langle \delta C_\ell^2\rangle \equiv
\int dC_\ell~C_\ell^2p(C_\ell|T) - \left[\int dC_\ell~C_\ell
				     p(C_\ell|T)\right]^2
\equiv \frac{2(C_\ell+N_\ell)^2}{(2\ell+1)f_{{\rm sky},\ell}^2},
\end{eqnarray}
where $N_\ell$ shows the contribution from instrumental noise and
a parameter $f_{{\rm sky},\ell}$ may be regarded as the effective
fraction of 
sky used for the analysis at each $\ell$. We know how to calculate
$N_\ell$ from the known properties of noise and beam transfer functions;
thus, for a given value of $C_\ell$, the only unknown quantity is
$f_{{\rm sky},\ell}$. Equation~(\ref{eq:variance}) thus provides 
definition of $f_{{\rm sky},\ell}$, which is a slowly-varying function
of $\ell$ (section~2.2.1 of \citep{verde/etal:2003}). 
This equation shows that there is an irreducible uncertainty even in the
absence of noise, $2C_\ell^2/[(2\ell+1)f_{{\rm sky},\ell}^2]$. This is
the so-called ``cosmic variance'' term, which arises from the fact that
$C_\ell$ is variance of CMB temperatures, and only $2\ell+1$
samples are available for estimating variance at each $\ell$.

\subsection{Polarization}

The polarization analysis is similar to the temperature analysis. We
begin with a Gaussian PDF for temperature and polarization:
\begin{equation}
 p(m) = \frac{\exp\left[-\frac12\sum_{ij}\sum_{ab}
			       m_i^{(a)}(S+N)^{-1}_{iajb}
			       m_j^{(b)}\right]}{\sqrt{\det[2\pi(S+N)]}},
\label{eq:polpdf}
\end{equation}
where $m=(\delta T,Q,U)$, and the signal matrix contains all the power
spectrum combinations such as $C_\ell^{TT}$, $C_\ell^{TE}$,
$C_\ell^{EE}$, and $C_\ell^{BB}$ (as well as parity-violating
combinations, $C_\ell^{TB}$ and $C_\ell^{EB}$, if necessary).
The explicit expressions are given in appendix of
\citep{katayama/komatsu:2011}.

\begin{figure}[t]
\centering\includegraphics[width=0.9\textwidth]{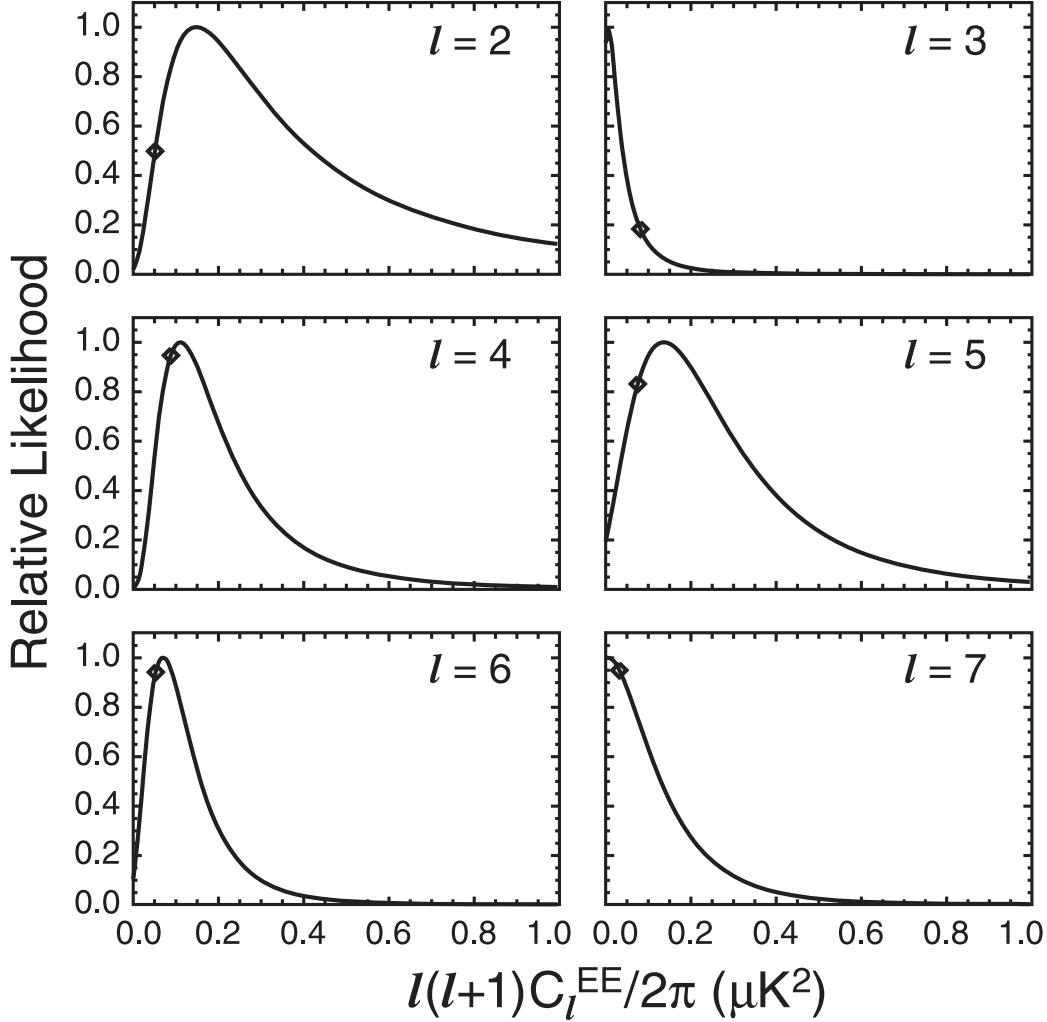}
\caption{Likelihood functions of the nine-year EE power
 spectrum for $\ell=2$ through 7 obtained from
 equation~(\ref{eq:polpdf}) (adapted from
 \citep{bennett/etal:2013}). These data fix the
 optical depth, $\tau$. The diamonds show the theoretical values of 
 the best-fit $\Lambda$CDM cosmological model. }
\label{fig:clee}
\end{figure}

The TE power spectrum does not add much to the parameter constraints
but is included in the model fits. The most important information we
obtain from the polarization likelihood is the optical depth, $\tau$,
from the EE power spectrum at $\ell\lesssim 10$. We can evaluate the
exact likelihood function given by equation~(\ref{eq:polpdf}) for such low
multipoles, using the steps described in appendix D of
\citep{page/etal:2007}. More precisely, we use equation~(\ref{eq:polpdf}) to
calculate the likelihood using the data at $\ell\le 23$. We use the
polarization maps at 33, 41, 61, and 94~GHz, while we use the ILC map
for the temperature. Figure~\ref{fig:clee} shows the
likelihood of the EE power spectrum,
$\ell(\ell+1)C_\ell^{EE}/(2\pi)$, for $\ell=2$ through 7.

It is still useful to compute the power spectrum of TE at high
multipoles. For this we use a simplified approach: we do not weight the 
temperature maps at 61 and 94~GHz, while we weight the 
polarization maps at 41, 61, and 94~GHz by $\sigma_0^2/n_{{\rm obs},i}$.
We then compute $(2\ell+1)^{-1}\sum_m a_{\ell m}^{T}a_{\ell m}^{E*}$,
and deconvolve the effects of the mask and weight following appendix A
of \citep{kogut/etal:2003}.

\begin{figure}[t]
\centering\includegraphics[width=0.9\textwidth]{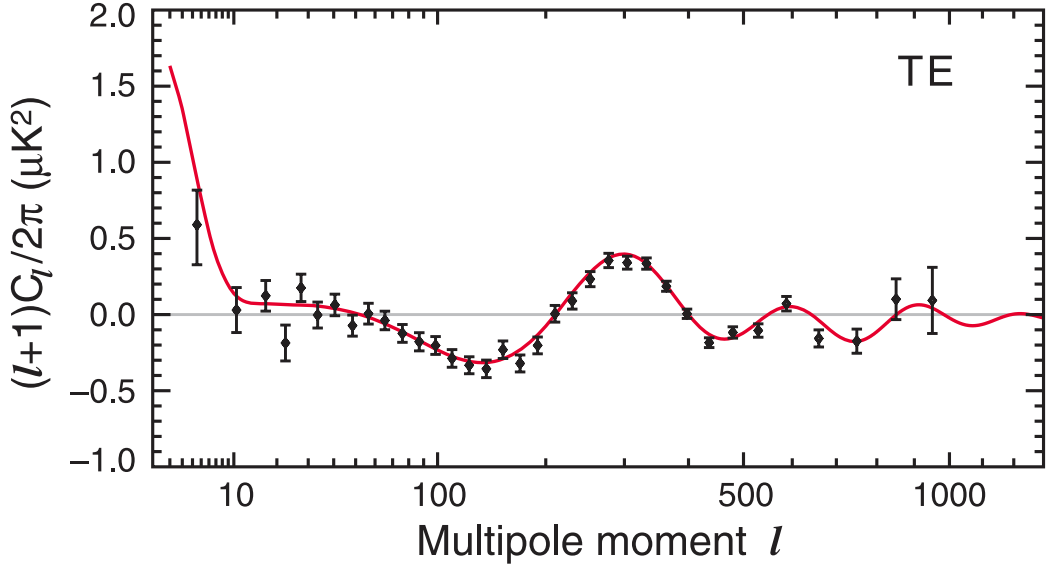}
\caption{Nine-year angular cross power spectrum of the CMB temperature
 and E-mode polarization
 (adapted from \citep{bennett/etal:2013}). While we measure $C^{TE}_\ell$ at
 each $\ell$ in $2\le \ell\le 1000$, the points with error bars show
 the binned values of $C_\ell^{TE}$ for clarity. The error bars show the 
 standard deviation of $C_\ell^{TE}$, which include both the
 instrumental noise and the cosmic variance. The solid line
 shows the theoretical curve of the best-fit $\Lambda$CDM cosmological model.}
\label{fig:clte}
\end{figure}

Figure~\ref{fig:clte} shows the nine-year measurements of $C^{TE}_\ell$ along
with estimates of the 68\%~CL error bars. The error bars are calculated
as follows:
\begin{eqnarray}
\label{eq:varianceTE}
\langle (\delta C^{TE}_\ell)^2\rangle =
 \frac{(C^{TT}_\ell+N^{TT}_\ell)(C^{EE}_\ell+N^{EE}_\ell)+(C_\ell^{TE})^2}{(2\ell+1)f^T_{{\rm
 sky},\ell}f^E_{{\rm sky},\ell}}, 
\end{eqnarray}
where $N_\ell^{TT}$ and $N_\ell^{EE}$ are the noise bias spectra of the
temperature and E-mode polarization, respectively, and 
$f^T_{{\rm  sky},\ell}$ and $f^E_{{\rm sky},\ell}$ are the effective sky
fractions of the temperature and E-mode polarization data, respectively.

\begin{figure}
\centering\includegraphics[angle=270,width=0.8\textwidth]{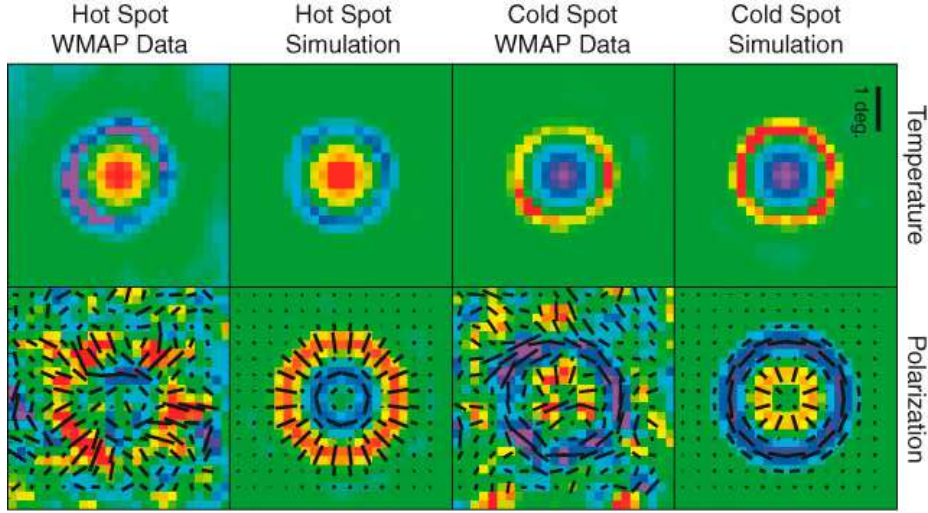}
\caption{Average images of temperature and polarization data. 12387 hot
 spots and 
 12628 cold spots are found in the {\sl WMAP} seven-year temperature
 maps, and the average images of hot and cold spots are shown in the top
 panels along with the corresponding simulated images. The bottom panels
 show the average images of the polarization maps around the locations
 of hot and cold temperature spots, as well as the corresponding simulated
 images. The size of each image is $5^\circ$ by $5^\circ$.
 The lines show the polarization directions, and their lengths are
 proportional to the magnitude of polarization. The colors of the
 polarization images are chosen such that blue and red show the
 tangential and radial polarization patterns, respectively.
 The data show the predicted tangential and radial polarization
 patterns ($E$-mode polarization), in excellent agreement with the
 predictions. The maximum of radial polarization around hot spots occurs
 at 1.2 degrees  from the center, whereas the maximum of tangential
 polarization around hot spots occurs at 0.6 degrees from the center.
 Figure adapted from \protect\url{http://wmap.gsfc.nasa.gov/media/101079/index.html} (Credit: NASA/WMAP Science Team).
}
\label{fig:pol}
\end{figure}

As the temperature and E-mode polarization are correlated, we can create
images of E-mode polarization around temperature spots by averaging the
polarization data around hot and cold spots. Figure~\ref{fig:pol} shows
average images of temperature and polarization data. We find that the
polarization data around hot and cold spots exhibit radial and
tangential polarization patterns, as predicted by simulations. What is
the physics behind them?

First of all, the necessary and sufficient conditions for generating
non-zero polarization of the CMB are to have Thomson scattering and
quadrupolar temperature anisotropy around an electron. Frequent Thomson
scatterings between photons and electrons suppress quadrupolar
temperature anisotropy around an electron, and thus we need to wait
until the photon decoupling epoch (at which photons and electrons become
less strongly coupled) to produce polarization. How is then quadrupolar
anisotropy around an electron created?

It turns out that polarization (for scalar modes) traces a velocity
gradient field of the plasma 
around gravitational potentials. Suppose that a packet of the plasma is
falling into the bottom of the potential well. Due to acceleration, a velocity
gradient is generated: the front of the packet falls faster than the
back of the packet. Therefore, an electron at the center of the packet
observes redshifted photons from both the front and back of the packet,
whereas there is no redshift or blueshift from the sides of the
packet. This produces a quadrupolar radiation pattern (colder along 
the motion of the packet and hotter in the perpendicular directions), and
the produced polarization is parallel to the motion of the packet. 
The polarization pattern around a spherically symmetric gravitational
potential well is {\it 
radial}, and the magnitude of radial polarization is maximal at twice
the sound horizon radius at the decoupling epoch  (or
1.2 degrees in the sky) from the bottom of the potential well
\citep{komatsu/etal:2011}. As the packet approaches the bottom of the
potential well, the packet decelerates because of a pressure
gradient. In the adiabatic initial condition, the photon density is high
at the bottom of the potential well, producing a pressure gradient to
decelerate motion of the plasma falling into the potential
well. The front of the packet falls slower than the
back of the packet. Therefore, an electron at the center of the packet
observes blueshifted photons from both the front and back of the packet,
whereas there is no redshift or blueshift from the sides of the
packet. This produces the opposite quadrupolar radiation pattern (hotter
along  
the motion of the packet and colder in the perpendicular directions), and
the produced polarization is {\it tangential} to the motion of the
packet. The magnitude of tangential polarization is maximal at the sound
horizon radius (or 
0.6 degrees in the sky) from the bottom of the potential well
\citep{komatsu/etal:2011}. 

These predictions have been confirmed by the {\sl WMAP} polarization
data. The bottom panels of Figure~\ref{fig:pol} show the average
polarization directions measured around hot and cold temperature spots. On these
angular scales (a few degrees), hot and cold spots correspond
to potential wells and 
hills, respectively. (The high photon energy density at the bottom of
the well overcomes the Sachs--Wolfe effect, turning potential wells into
hot spots in the sky.) Therefore, we expect each hot spot to come with the
radial and tangential polarization patterns at 1.2 and 0.6 degrees from
the center, respectively, and each cold spot to come with the opposite
patterns. As the magnitude of polarization is small, {\sl WMAP}
cannot detect polarization around each spot; however, by averaging
polarization patterns around many spots, we can detect
polarization. There are 12387 hot spots and  12628 cold spots outside
the Galactic mask in the {\sl WMAP} seven-year temperature
map. Averaging the polarization data around these spots, the expected
polarization patterns (shown in the ``Simulation'' columns in
Figure~\ref{fig:pol}) are clearly detected in the data  (shown in the
``WMAP Data'' columns) \citep{komatsu/etal:2011}. 

The TE cross power spectrum and the average polarization images offer a
 powerful, precision test of the standard cosmological model. We fix
 the basic 
 six cosmological parameters by fitting the temperature power spectrum
 at $2\le\ell\le 1200$ and the $E$-mode polarization power spectrum at
 low multipoles.  We can then predict the cross
 power spectrum {\it without any more additional free parameters}. The
 prediction matches with the data at the precision shown in
 Figure~\ref{fig:clte} and \ref{fig:pol}. This is a great triumph of the
 standard cosmological model.

The TE cross power spectrum  offers also a powerful test of one of the
generic predictions of cosmic inflation: the presence of
``super-horizon'' fluctuations, whose wavelength is greater than the
horizon size at the decoupling time. This test is possible because
polarization is generated only when there are free electrons. The
reionization of the universe at $z\lesssim 10$ can generate polarization
only on very large angular scales, $\ell\lesssim 10$; thus, any TE
signals at high multipoles must be generated at the decoupling
epoch. The angle that subtends the radius of the horizon at the
decoupling epoch is 1.2 degrees, which corresponds to
$\ell=150$. Therefore, the anti-correlation seen in the TE cross power
spectrum at $\ell<150$ provides the direct evidence for the presence of
super-horizon fluctuations at the decoupling epoch, i.e., the key
prediction of inflation \citep{spergel/zaldarriaga:1997,peiris/etal:2003}.

\section{Cosmological Parameters from Power Spectra}

\subsection{Standard six parameters}

\begin{table}
\caption{Six cosmological parameters of the standard flat $\Lambda$CDM
 model, determined from the CMB data alone. We show the constraints from
 the {\sl WMAP} nine-year temperature and polarization power spectra
 \citep{bennett/etal:2013,hinshaw/etal:2013}; 
 the {\sl WMAP} data combined with the  {\sl ACT} \citep{das/etal:2011}
 and {\sl SPT} \citep{keisler/etal:2011} temperature spectra (``{\sl
 WMAP}+{\sl ACT}+{\sl SPT}''); and  the {\sl Planck} 
 15.5-month  temperature power spectrum combined with the {\sl WMAP} nine-year
 polarization power spectrum (``{\sl Planck}+WP'') \citep{planck:16}.} 
\centering\begin{tabular}{@{}c@{\hspace{25pt}}ccc@{}}
\hline\hline
Parameters & {\sl WMAP} only & {\sl WMAP}+{\sl ACT}+{\sl SPT} & {\sl Planck}+WP \\
\hline
$\Omega_bh^2$ &$0.02264\pm 0.00050$ & $0.02229\pm 0.00037$ & $0.02205\pm 0.00028$ \\
             %$0.02217\pm 0.00033$ \\
$\Omega_ch^2$ &$0.1138\pm 0.0045$ & $0.1126\pm 0.0035$ & $0.1199\pm 0.0027$ \\ %& $0.1186\pm 0.0031$ \\
$\Omega_\Lambda$ & $0.721\pm 0.025$ & $0.728 \pm 0.019$ & $0.685^{+0.018}_{-0.016}$  \\ %& $0.693\pm 0.019$ \\
$10^9\Delta^2_{\mathcal R}$ & $2.41\pm 0.10$ & $2.167\pm 0.056$ &
         $2.196^{+0.051}_{-0.060}$ \\ %& $2.19^{+0.12}_{-0.14}$ \\
$n_s$ & $0.972\pm 0.013$ & $0.9646\pm 0.0098$ & $0.9603\pm 0.0073$ \\ %& $0.9635\pm 0.0094$ \\
$\tau$ & $0.089\pm 0.014$ & $0.084\pm 0.013$ & $0.089^{+0.012}_{-0.014}$ \\ %& $0.089\pm 0.032$ \\
\hline\hline
\end{tabular}
\label{tab:parameters}
\end{table}

The {\sl WMAP} nine-year temperature and polarization data are
consistent with the minimal six-parameter flat $\Lambda$CDM model
\citep{bennett/etal:2013,hinshaw/etal:2013}. 
The high-$\ell$ temperature
power spectrum ($33\le \ell\le 1200$) gives $\chi^2=1200$ for 1168
degrees of freedom, with the probability to exceed (PTE) of 25.1\%. The
high-$\ell$ TE spectrum ($24\le \ell \le 800$) gives $\chi^2=815.4$ for 777
degrees of freedom, with PTE of 16.5\%. Therefore, the best-fit
$\Lambda$CDM model is a good fit to the high-$\ell$ data.

While the low-$\ell$ temperature likelihood ($2\le \ell\le 32$) based on the
BR estimator does not give a $\chi^2$ value, we find that the best-fit
$\Lambda$CDM model is consistent with the distribution of spectra
generated by the BR estimator.
The low-$\ell$ polarization likelihood
($2\le \ell\le 23$) evaluated directly in pixel space gives
$\chi^2=1321$ for 1170 degrees of freedom with PTE of 0.13\%, which is
unusually low. This excess $\chi^2$ can be interpreted as an additional
noise component (due to, for example, residual foreground emission) of
$0.27~{\mu}{\rm K}$ per $N_{\rm side}=8$ pixel 
($7.3^\circ$ on a side), which is significantly lower than the average
standard deviation of $0.86\pm 0.17~{\mu}{\rm K}$. 
We confirm that this
excess noise does not affect the determination of $\tau$ by using
differences bewtween frequencies. See section~7.1 of
\cite{bennett/etal:2013} for details.

We assume flat priors on the following six parameters: the amplitude of the
primordial power spectrum at $k=0.002~{\rm Mpc}^{-1}$, $\Delta_{\mathcal
R}^2$, the tilt of the primordial power spectrum, $n_s$, the physical baryon
density parameter, $\Omega_b h^2$, the physical CDM density parameter, $\Omega_c
h^2$, the cosmological constant density parameter, $\Omega_\Lambda$,
and the optical depth of the reionization, $\tau$.

Table~\ref{tab:parameters} summarizes the constraints of the
cosmological parameters from the CMB data alone. Adding the
smaller-scale CMB data from the {\sl Atacama Cosmology Telescope} ({\sl
ACT}) \citep{das/etal:2011} and the {\sl South Pole Telescope} ({\sl
SPT}) improves the parameter constraints significantly. In particular,
statistical significance of a deviation of $n_s$ from unity increases
from $2.1\sigma$ to $3.6\sigma$. With additional cosmological measurements 
(Baryon Acoustic Oscillation
\citep{beutler/etal:2011,blake/etal:2012,padmanabhan/etal:2012,anderson/etal:2013}
and the local Hubble constant \citep{riess/etal:2011}) this 
improves to $n_s=0.9608\pm 0.0080$ (68\%~CL), a $4.9\sigma$  
deviation from unity. This is a great achievement in
cosmology, providing strong evidence for cosmic inflation.

The parameters found from the {\sl
Planck} 15.5-month data combined with the {\sl WMAP} low-$\ell$
polarization data (``{\sl Planck}+WP'') are consistent with the {\sl
WMAP} and {\sl WMAP}+{\sl ACT}+{\sl SPT} parameters to within the quoted
error bars. With the {\sl Planck}+WP combination, $1-n_s$ is 
detected at $5.4\sigma$. This is the first time that $n_s<1$ is detected
with $>5\sigma$ from the CMB data alone. 

The {\sl WMAP} measurements also provide definitive evidence for the
existence of non-baryonic dark matter with $\Omega_c/\Omega_b=5.0\pm
0.2$ (68\%~CL). This
measurement comes from a combination of the ratio of the heights of odd
and even acoustic peaks giving $\Omega_bh^2$ and the ratio of the
heights of the first and other peaks giving the total matter density
contributing to gravitational potential well. In other words, {\sl
WMAP} measures the density of matter which interacts with photons, and
which does not. (No matter is left behind.) The difference between the
two provides definitive evidence for non-baryonic dark matter.

{\sl WMAP} has erased lingering doubts about the existence of dark
energy. This measurement comes from the peak positions giving the
angular diameter distance, $d_A=c\int_0^{z_*} dz/H(z)$, where $z_*=1091$
is the redshift of the photon decoupling. The Friedmann equation relates
the Hubble expansion rate, $H(z)$, to the total energy density in the
universe. As the integral is dominated by low redshift contributions,
this provides an estimate of the total energy density in a local
universe. As we have the complete account of matter density in the
universe at any redshifts after $z_*$, the difference between the total
energy density inferred from $d_A$ and the total matter density gives the energy
density of some substance which is {\it not even matter}, i.e., dark
energy. Strictly speaking, this measurement is possible if we assume
flatness of the universe or combine the {\sl WMAP} data with other cosmological 
data (such as the local Hubble constant measurements). Alternatively, we
can use the effects of gravitational lensing on the CMB to detect dark energy
from the CMB data alone \citep{sherwin/etal:2011}.

\subsection{Parameters beyond flat $\Lambda$CDM}

The minimal six-parameter model fits all the data we have at the moment
(with a possible exception of the tensor-to-scalar ratio, $r$, which the
BICEP/Keck Array collaboration claims to have found recently from the
B-mode polarization at degree angular scales \citep{bicep2:2014}).
As a result, the {\sl WMAP} data (sometimes in combination with 
other CMB and non-CMB data) place stringent limits on the parameters
beyond the minimal model. 

Spatial geometry of the universe is consistent with flat (Euclidean) space.
By combining the {\sl WMAP}+{\sl ACT}+{\sl SPT} with the CMB lensing
data, we find $\Omega_k=-0.001\pm 0.012$ (68\%~CL). When other non-CMB
data (Baryon Acoustic Oscillation and the local Hubble constant
measurements) are added, we find a stringent limit of
$\Omega_k=-0.0027^{+0.0039}_{-0.0038}$ (68\%~CL), i.e., 0.4\% measurement.

Dark energy is consistent with a cosmological constant. The constraints
on the equation of state parameter, $w$, are consistent with $w=-1$
typically to within 10\% (95\%~CL), depending on the data combinations.

The cosmic neutrino background affects temperature anisotropy of the CMB
in four ways: peak locations, early integrated Sachs-Wolfe effect,
anisotropic stress, and enhanced damping tail (see section 4.3.1 of
\citep{hinshaw/etal:2013} for summary). Using these effects and the {\sl
WMAP} five-year data, we have made the first (indirect) detection of the
cosmic neutrino background \citep{komatsu/etal:2009}. The CMB data give
the total energy density of neutrinos, $\rho_\nu=(7\pi^2/120)N_{\rm
eff}T_\nu^4$, where $N_{\rm eff}$ is the effective number of neutrino species.
 Assuming the
standard thermal history of the universe relating the asymptotic neutrino
temperature to the CMB temperature as
$T_\nu=(4/11)^{1/3}T_{\rm cmb}$, we use the nine-year data combined with
{\sl ACT} and {\sl SPT} to find $N_{\rm eff}=3.89\pm 0.67$ (68\%~CL),
consistent with the standard value of 3.046 to within $2\sigma$.

The damping tail of the CMB is sensitive to the primordial helium
abundance, $Y_{\rm He}$. The more helium we have, the more electrons are
captured by helium nuclei before the decoupling, the fewer electrons are
available at the decoupling, the more diffusion damping results.
We have made the first detection of this effect
by combining the {\sl WMAP} seven-year data and the small-scale CMB data
\citep{komatsu/etal:2011}. The nine-year data combined with {\sl ACT} and
{\sl SPT} give $Y_{\rm He}=0.299\pm 0.027$ (68\%~CL), consistent with
the standard value of 0.25 to within $2\sigma$.
These measurements of $N_{\rm eff}$ and $Y_{\rm He}$ offer a unique test
of the Big Bang nucleosynthesis \citep{steigman:2012}. Our measurements
are consistent with the standard Big Bang nucleosynthesis calculations.

The {\sl WMAP} nine-year data alone place a limit on the sum of neutrino masses,
$\sum m_\nu<1.3$~eV (95\%~CL). Adding the Baryon Acoustic Oscillation
and the local Hubble constant measurements, the limit improves to $\sum
m_\nu<0.44$~eV (95\%~CL). 

Single-field inflation models predict that the fluctuations in matter
and photons trace each other, obeying the adiabatic relation of
$\delta\rho_m/\rho_m=(3/4)\delta\rho_\gamma/\rho_\gamma$. We find that
this relation holds to better than 7\% (95\%~CL). This limit plays an
important role in constraining the parameter space of axion dark matter
models \citep{komatsu/etal:2009,komatsu/etal:2011}.

The shape of the primordial power spectrum is sensitive to the physics
of inflation. The ``running spectral index,'' $dn_s/d\ln k$, is
typically predicted to be of order $(n_s-1)^2={\mathcal O}(10^{-3})$.
The {\sl WMAP} nine-year data alone give $dn_s/d\ln k=-0.019\pm 0.025$,
while adding the small-scale CMB data improves the limit to $dn_s/d\ln
k=-0.022^{+0.012}_{-0.011}$ (68\%~CL), consistent with a power-law power
spectrum to within $2\sigma$.

Finally, inflation generates nearly scale-invariant tensor mode metric
perturbations (gravitational waves) \citep{starobinsky:1979}, $h_{ij}$,
which also contribute to the 
observed temperature and polarization anisotropies of the CMB. The
amplitude of $h_{ij}$ is parametrized by the ``tensor-to-scalar ratio,''
$r$, defined by $r\equiv 2\langle h_{ij}h^{ij*}\rangle/\langle|{\mathcal
R}|^2\rangle$. The {\sl WMAP} data alone give $r<0.38$ (95\%~CL), which
improves to $r<0.17$ by 
adding the small-scale CMB data. Adding the Baryon Acoustic Oscillation
measurements improves the limit further to
$r<0.12$ (95\%~CL). This limit largely comes from the low multipole
temperature data (see section 3.2.3 of \citep{komatsu/etal:2009}), and
thus it is sensitive to our assumption of a 
power-law power spectrum. Including the running index relaxes the limits
to $r<0.43$ (95\%~CL) with a large running index, $dn_s/d\ln k\approx
-0.04$, more or less independent of the data sets used. 

{\sl WMAP} did not have sufficient sensitivity to detect B-mode
polarization. Recently the BICEP/Keck Array collaboration claimed to
have found B-mode polarization at degree angular scales at 150~GHz. If this
signal is cosmological and originates from gravitational waves from
inflation, it corresponds to $r\approx 0.1-0.2$  \citep{bicep2:2014}. As
the measurement was done only at one frequency (the BICEP1 data at
100~GHz are too noisy to be useful), confirmation of the signal at other
frequencies must be made to reject foregrounds. In any case, an independent 
detection from an independent group is required before interpreting the 
detected signal as inflationary. 

\section{Tests of Gaussianity with Angular Bispectrum}
\label{sec:ng}
Inflation predicts that primordial fluctuations originate from quantum
fluctuations, and the distribution of primordial fluctuations is
nearly a Gaussian distribution (see \citep{bartolo/etal:2004} for a
review). Sustained inflationary expansion for at least 50 $e$-folds
requires a field driving inflation to be weakly coupled. The wave
function of quantum fluctuations of a scalar field with no interaction
in the ground state is 
precisely a Gaussian; thus, a weakly coupled field is nearly a Gaussian field.
The linear physics preserves Gaussianity, and thus CMB
temperature and polarization anisotropies are predicted to obey Gaussian
statistics with high precision. Confirmation of this prediction gives
strong evidence for the quantum origin of primordial fluctuations.

When the distribution is not a Gaussian, the PDF is no longer given by
equation~(\ref{eq:pdf}). However, when a departure from Gaussianity,
i.e., non-Gaussianity, is small, we may approximate the PDF by ``Taylor-expanding'' around a Gaussian distribution. Let us do this in harmonic space. We
obtain an expanded PDF for the spherical harmonics coefficients as
\citep{amendola:1996,taylor/watts:2001} 
\begin{eqnarray}
\nonumber
 p(a)&=&
\left[1-\frac16\sum_{{\rm all}~\ell_im_j}
\langle a_{\ell_1m_1}a_{\ell_2m_2}a_{\ell_3m_3}\rangle
\frac{\partial}{\partial a_{\ell_1m_1}}
\frac{\partial}{\partial a_{\ell_2m_2}}
\frac{\partial}{\partial a_{\ell_3m_3}}
\right]\\
& &\times
 \frac{e^{-\frac12\sum_{\ell m}\sum_{\ell'm'}a^*_{\ell m}(C^{-1})_{\ell m,\ell'm'}a_{\ell'm'}}}{\sqrt{\det(2\pi C)}},
\end{eqnarray}
where $C_{\ell m,\ell'm'}\equiv \sum_{ij}Y_{\ell m,i}(S+N)_{ij}Y_{\ell' m',j}^*$ is the signal plus noise covariance matrix in
harmonic space. (We do not write indices for DAs or years for
simplicity.) Here, the expansion is truncated at the three-point function
(bispectrum) of $a_{\ell m}$, and thus we have assumed that the
connected four-point and higher-order correlation 
functions are negligible compared to the power spectrum and
bispectrum. (This condition is not always satisfied.)

By evaluating the above derivatives, we obtain\footnote{Babich
\citep{babich:2005} derived this formula for
$C_{\ell m,\ell'm'}=C_{\ell}\delta_{\ell \ell'}\delta_{mm'}$.} 
\begin{eqnarray}
\nonumber
& &p(a)=
\frac1{\sqrt{\det(2\pi C)}}
\exp\left[-\frac12\sum_{\ell m}\sum_{\ell'm'}a_{\ell m}^*(C^{-1})_{\ell
     m,\ell'm'}a_{\ell'm'}\right]\\   
\nonumber
&\times&
\left\{
1+\frac16\sum_{{\rm all}~\ell_im_j}
\langle a_{\ell_1m_1}a_{\ell_2m_2}a_{\ell_3m_3}\rangle
\left[
(C^{-1}a)_{\ell_1m_1}(C^{-1}a)_{\ell_2m_2}(C^{-1}a)_{\ell_3m_3}
\right.\right.\\
& &
\left.\left.
-3(C^{-1})_{\ell_1m_1,\ell_2m_2}(C^{-1}a)_{\ell_3m_3}
\right]\right\}.
\label{eq:ngpdf}
\end{eqnarray}
This formula is useful, as it tells us how to estimate the {\it angular bispectrum},
$\langle a_{\ell_1m_1}a_{\ell_2m_2}a_{\ell_3m_3}\rangle$, optimally from given
data by maximizing this PDF. In practice,
we usually parametrize the bispectrum using a few parameters (e.g., $f_{\rm
NL}$), and estimate those parameters from the data by maximizing the PDF
with respect to the parameters.

In the limit that the contribution of the connected four-point function
(trispectrum) to the PDF is 
negligible compared to those of the power spectrum and bispectrum,
equation~(\ref{eq:ngpdf}) contains all the information on
non-Gaussian fluctuations characterized by the covariance matrix,
$C_{\ell_1m_1,\ell_2m_2}=\langle a_{\ell_1m_1}^*a_{\ell_2m_2}\rangle$, and the angular
bispectrum, $\langle a_{\ell_1m_1}a_{\ell_2m_2}a_{\ell_3m_3}\rangle$.
This approach can be extended straightforwardly to the trispectrum if
necessary. 

As the bispectrum has three angular wavenumbers, $\ell_1$, $\ell_2$ and
$\ell_3$, it can form triangles with various shapes. Among all the
shapes, the so-called ``local-form bispectrum,'' parametrized by a
non-linear parameter $f_{\rm NL}$ \citep{komatsu/spergel:2001}, carries a special 
significance, as detection of a large local-form bispectrum would rule
out all inflation models based on a single energy component with a
Bunch-Davies initial vacuum state and an attractor solution (see
\citep{chen/etal:2013} for the latest discussion on this theorem).
This triangle has the largest amplitude in the ``squeezed
configurations'' in which one of the wavenumbers, say $\ell_3$, is much
smaller than the other two, i.e., $\ell_3\ll \ell_1\approx \ell_2$
\citep{babich/creminelli/zaldarriaga:2004}. 
Detailed descriptions on what this bispectrum is and what the other
shapes are, as well as on how to measure them can be found in
\citep{komatsu:2010}.

Using the foreground-reduced {\sl WMAP} nine-year temperature data at 61
and 94~GHz with the KQ75 mask, we find $f_{\rm NL}=37\pm 20$
(68\%~CL), which is consistent with zero to within $2\sigma$; thus, the
measurement agrees with the basic prediction of single-field inflation
models with a Bunch-Davies initial vacuum state and an attractor
solution. The {\sl Planck} improves this limit greatly by finding
$f_{\rm NL}=2.7\pm 5.8$ (68\%~CL) \citep{planck:24}. 

One way to generate the local-form bispectrum is to write the primordial
curvature perturbation as ${\mathcal R}({\bf x})={\mathcal R}_L({\bf
x})+\frac35f_{\rm NL}{\mathcal R}^2_L({\bf x})$. This form is called the ``local
form'' because both sides are evaluated at the same spatial location,
${\bf x}$. Here, ${\mathcal R}_L$ is a
 Gaussian random field, and the curvature perturbation is defined such
 that the linear Sachs--Wolfe effect gives $\delta T/T=-{\mathcal
 R}_L/5$. Using this form and the fact that the variance of ${\mathcal R}$
 is $2\times 10^{-9}$, we find that the 95\% upper bound from
{\sl Planck}, $f_{\rm NL}<14$, implies that the observed CMB is Gaussian
to the precision of 0.04\% or better. This is a remarkable degree of
Gaussianity, which provides strong evidence that the observed CMB
fluctuations originate from quantum fluctuations generated during
single-field inflation.

\section{Implications for  inflation}

Models of inflation
\citep{starobinsky:1980,sato:1981,guth:1981,linde:1982,albrecht/steinhardt:1982}
make specific, testable predictions. The simplest models based upon a
single energy component (scalar field), which slowly rolls down on its
potential and drives a sustained quasi-exponential expansion for at
least 50 $e$-folds, predict that the observable universe is homogeneous
and isotropic with flat geometry, and is filled with small fluctuations
which are precisely adiabatic and nearly Gaussian (before fluctuations
become non-linear). Both scalar and tensor fluctuations with various 
wavelengths are generated during inflation. The wavelengths of these
fluctuations can exceed the horizon size at the decoupling epoch,
and the amplitude of these fluctuations weakly depends on wavelengths. 

{\it All} of these predictions fit the {\sl WMAP} data remarkably well:
flatness is measured with 0.4\% precision (from {\sl WMAP}
combined with the Baryon Acoustic Oscillation and the local Hubble constant
measurements); the adiabatic condition holds to better than 7\%
precision; a deviation from Gaussian fluctuations is restricted to be
less than 0.2\% (and 0.04\% with the {\sl Planck} 2013 data); and the
presence of super-horizon fluctuations at decoupling is
decisively detected in the TE cross power spectrum at $\ell<150$.
The {\sl WMAP} data combined with the Baryon Acoustic Oscillation and the local
Hubble constant measurements find convincing evidence for the scale
dependence of the scalar initial power spectrum with 4.9$\sigma$
significance, with the best-fit value in agreement with the first
prediction made in \citep{mukhanov/chibisov:1981}.

\begin{figure}[t]
\centering\includegraphics[width=0.9\textwidth]{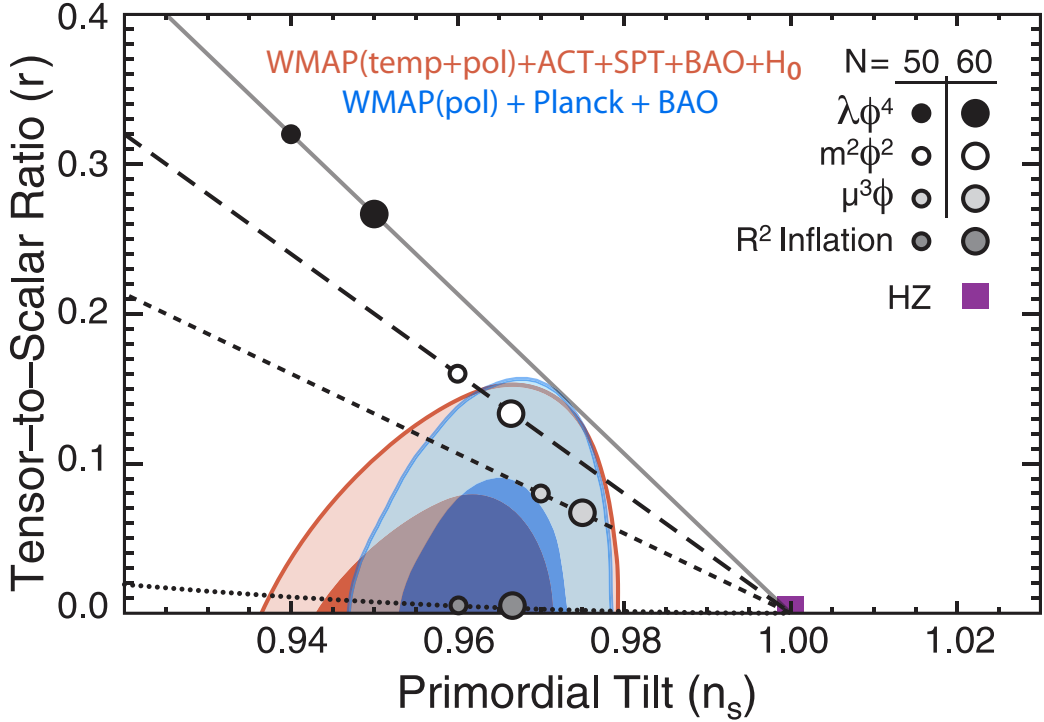}
\caption{Two-dimensional joint marginalized constraints (68\% and
 95\%~CL) on the primordial tilt, $n_s$, and the tensor-to-scalar ratio,
 $r$. The red contours show the constraint from the {\sl WMAP} nine-year
 data combined with  the small-scale CMB temperature data ({\sl ACT} and
 {\sl SPT}), the Baryon Acoustic Oscillation data, and the local Hubble
 constant. The blue contours show the 
 constraint from the  {\sl Planck} 15.5-month temperature data combined
 with the {\sl WMAP} nine-year polarization data and the Baryon Acoustic
 Oscillation data. The symbols show the predictions from single-field
 inflation with monomial potentials, $V(\phi)\propto \phi^n$
 \cite{linde:1983}, with $n=4$  (black), $2$ (white), and $1$ (light
 grey), and with a $R^2$ term in the gravitational action (dark grey)
 \cite{starobinsky:1980}.}
\label{fig:ns-r}
\end{figure}

While {\sl WMAP} did not find signatures of tensor fluctuations, the
upper bound on the tensor-to-scalar ratio inferred 
from the temperature data is consistent with many single-field inflation
models. Figure~\ref{fig:ns-r} compares the limits on $n_s$ and $r$ with
a few representative single-field inflation models.

\section{Conclusion}

The nine years of observations of the {\sl WMAP} satellite have taught
us many things. The current universe is 13.77 billion years old, and
consists of 4.6\% atoms, 24\% cold dark matter, and 71\% dark energy \cite{bennett/etal:2013,hinshaw/etal:2013}. The
nature of dark energy is consistent with that of a cosmological
constant. The spatial geometry of the universe is consistent with
Euclidean geometry. The universe is filled with neutrinos, whose
abundance is consistent with the standard model of particle physics. The
mass of neutrinos is much less than 1~eV. 

The measured properties of primordial fluctuations
such as adiabaticity, Gaussianity, and near scale invariance all point
toward a remarkable scenario: the observed fluctuations originate from
quantum fluctuations generated during inflation driven by a single
energy component. {\sl WMAP} offered a number of stringent tests of the
simplest inflation scenarios: (1) flat universe, (2) adiabatic fluctuations, (3)
super-horizon fluctuations, (4) nearly, but not exactly, scale-invariant
initial power spectrum, and (5) Gaussian fluctuations. The simplest
scenarios passed all of these tests. The {\sl Planck} 2013 data have
confirmed all of these findings with greater precision. 

Yet, neither the {\sl WMAP} nor the {\sl Planck} 2013 data detect 
the signature of primordial gravitational waves from
inflation in CMB. Detecting and characterizing the 
B-mode polarization of the CMB is the next milestone in cosmology. While the
BICEP/Keck Array collaboration claims to have found the B-mode
polarization from inflationary gravitational waves at 150~GHz,
confirmation of the signal at other frequencies and with an independent
experiment must be made before we claim a victory in observing all of
the inflation predictions.

\section*{Acknowledgment}

The {\sl WMAP} mission was made possible by the support of the Science
Mission Directorate Office at NASA Headquarters. This research has made
use of NASA's Astrophysics Data System Bibliographic Services. 
All of the scientific products of the {\sl WMAP} mission are made
publicly available at the Legacy Archive for Microwave Background
Data Analysis (LAMBDA; \url{http://lambda.gsfc.nasa.gov/}).
Support for LAMBDA is 
provided by NASA Headquarters. We acknowledge use of the HEALPix
\citep{gorski/etal:2005}, CAMB \citep{lewis/challinor/lasenby:2000}, and
CMBFAST \citep{seljak/zaldarriaga:1996} packages. 
EK would like to thank the members of the {\sl WMAP} science team for
letting him join the team and for wonderful collaborations over the last
13 years. EK's work has been supported in part by an Alfred P. Sloan
Research Fellowship, NASA grants NNX08AL43G and NNX11AD25G, and NSF
grants AST-0807649 and PHY-0758153. CLB thanks the entire {\sl WMAP} Science 
Team for the many years of the close, careful, and productive collaboration  
that advanced cosmology in a significant manner.


%\bibliographystyle{ptephy}
%\bibliography{references}
%
% once the .bbl file has been generated then place the text in your article.

%\vfill\pagebreak

%\begin{thebibliography}{9}
%\end{thebibliography}

\end{document}